\documentclass[aps,pre,groupedaddress,superscriptaddress,twocolumn,showkeys,showpacs,amsmath,amssymb]{revtex4-1}
\usepackage{amsfonts}
\usepackage{bm}
\usepackage[colorlinks=true,
    linkcolor=blue,
    citecolor=blue,
    urlcolor=blue]{hyperref}

\begin{document}

\title{First principles approach to the Abraham-Minkowski controversy for the momentum of light in general linear non-dispersive media}

\author{Tom\'as Ramos}
\email{tomas.ramos@uibk.ac.at}
\affiliation{Institute for Theoretical Physics, University of Innsbruck, A-6020 Innsbruck, Austria and \\Institute for Quantum Optics and Quantum Information of the Austrian Academy of Sciences, A-6020 Innsbruck, Austria}

\author{Guillermo F. Rubilar}
\email{grubilar@udec.cl}
\affiliation{Departamento de F\'isica, Universidad de Concepci\'on, Casilla 160-C, Concepci\'on, Chile}

\author{Yuri N. Obukhov}
\email{obukhov@ibrae.ac.ru}
\affiliation{Theoretical Physics Laboratory, Nuclear Safety Institute, Russian Academy of Sciences, B.~Tulskaya 52, 115191 Moscow, Russia}

\begin{abstract}
We study the problem of the definition of the energy-momentum tensor of light in general moving non-dispersive media with linear constitutive law. Using the basic principles of classical field theory, we show that for the correct understanding of the problem, one needs to carefully distinguish situations when the material medium is modeled either as a background on which light propagates or as a dynamical part of the total system. In the former case, we prove that the (generalized) Belinfante-Rosenfeld (BR) tensor for the electromagnetic field coincides with the Minkowski tensor. We derive a complete set of balance equations for this open system and show that the symmetries of the background medium are directly related to the conservation of the Minkowski quantities. In particular, for isotropic media, the angular momentum of light is conserved despite of the fact that the Minkowski tensor is non-symmetric. For the closed system of light interacting with matter, we model the material medium as a relativistic non-dissipative fluid and we prove that it is always possible to express the total BR tensor of the closed system either in the Abraham or in the Minkowski separation. However, in the case of dynamical media, the balance equations have a particularly convenient form in terms of the Abraham tensor. Our results generalize previous attempts and provide a first principles basis for a unified understanding of the long-standing Abraham-Minkowski controversy without \emph{ad hoc} arguments.
\end{abstract}


\maketitle

\section{Introduction}

The proper definition and interpretation of the energy-momentum tensor of the electromagnetic field in material media is an old physical problem. In the literature, it is often refered to as the Abraham-Minkowski (AM) controversy which dates back to the beginning of the last century when Abraham \cite{abraham1,abraham2} postulated an \textit{ad hoc} symmetric energy-momentum tensor for the electromagnetic field in matter to replace the asymmetric Minkowski tensor \cite{minkowski}. Historically, the controversy has been mainly focused in studying light in simple media, for instance, isotropic and homogeneous media at rest, described uniquely by its refraction index $n$. In the simplest case of a plane wave propagating in these media, the Minkowski and Abraham momentum densities read \cite{milonniboyd,dielectricslab}
\begin{equation}
\bm{\pi}^M=n\frac{\cal{U}}{c}\hat{\bm{k}},\qquad \bm{\pi}^A=\frac{1}{n}\frac{\cal{U}}{c}\hat{\bm{k}},\label{rivals}
\end{equation} 
where $\cal{U}$ is the energy density of light inside the medium, $\hat{\bm{k}}$ the propagation direction unit vector and $c$ the velocity of light in vacuum.

The first attempts to understand the controversy, both theoretically \cite{pauli,wgordon,tamm1939,vonlaue,moller1,balazs,jpgordon,israel,brevik} and experimentally \cite{jonesrichards,james,ashkin,wlw,ww,jonesleslie,gibson,kristensen}, were mainly directed to find which of the two expressions (\ref{rivals}) was the correct one to describe the momentum of light in matter. The discussion brought more contradictory arguments and confusion than clarifications, delaying the understanding of the problem. A remarkable step towards resolution of the controversy was the theoretical work of Penfield and Haus in 1966 \cite{penfieldhaus1,penfieldhaus1967}, who treated the material medium interacting with the electromagnetic field as a dynamical part of the total system and used a variational approach to derive an expression for the total energy-momentum tensor of the closed system. Soon it was recognized that the Abraham and Minkowski tensors both provide valid decompositions of the same total tensor into field and matter parts \cite{robinson,mikura,maugin,kranys}, however this idea was somehow under-appreciated in the literature. In the past few years the discussion of the optical momentum in media has been revisited mainly due to the increasing interest in the study of optical forces in nanotechnology \cite{ashkintweesers,revolutiontweesers,binding} and metamaterials \cite{cloak0,cloak}. A large number of new theoretical papers on the AM controversy were written \cite{loudon1,obukhovhehl,padgettbarnettloudon,loudon,garrison,milonniboyd2,leonhardt,hindsbarnett,mansuripur5,mansuripur,
crenshaw,bradshawboydmilonni,shevchenko2,brevik2010,brevik2012,kempletter} and also new experiments were reported \cite{campbell,she,rikken,rikken2}, but most authors unfortunately did not take into account the ideas of Penfield and Haus and continued to look for the ``correct'' momentum of light in matter. Exceptions were the review by Pfeifer et al. \cite{pfeifer} in 2007 and our work in 2008 \cite{obukhov1}, where we developed a relativistic variational model for simple moving media using a Penfield-Haus type approach. In 2010 Barnett and Loudon \cite{barnett2010,barnettloudon2} also reanalysed the AM controversy in the spirit of Penfield and Haus and argued that both the Minkowski and Abraham momenta are correct, in the sense that both can be measured, but in different situations. They identify the Abraham momentum as the ``kinetic'' momentum of light in matter based mainly on the Einstein box thought experiment \cite{balazs,dielectricslab} and the Minkowski momentum as the ``canonical'' momentum because it is related to the generator of translations in the medium \emph{at rest}. This way of understanding the problem has been recently popularised as ``the final'' resolution of the AM controversy and the original ideas of Penfield and Haus have become more accepted \cite{milonniboyd,dielectricslab,saldanha,dodin,crenshaw2013,
crenshaw2014,webb,kempresolution,baxterloudon,griffiths}. For more historical details of the AM controversy, see the review \cite{pfeifer}; for a more accessible presentation of the work of Penfield and Haus, see \cite{obukhov1}.

In this work we develop a field-theoretical approach to provide a more fundamental understanding of the resolution of the AM controversy, and extend the earlier results to the case of the general linear anisotropic and moving non-dispersive material media. Our model represents a generalization of the theory of Penfield and Haus \cite{penfieldhaus1,penfieldhaus1967}, and it is a natural extension of our previous works \cite{obukhov1,obukhovhehl2007,ORR02}. We assume the validity of the macroscopic Maxwell equations with a general linear constitutive relation and use the principles of relativistic fluid dynamics to derive a complete Lagrangian model that describes the electromagnetic field in matter in two situations: (a) when the medium is a non-dynamical background for light and (b) when the medium is dynamically coupled to light forming a closed system. In each case from the corresponding field equations we derive the balance equations for the physically meaningful and predictive quantities. The use of the Minkowski or Abraham tensors for the description of light in material media is a matter of choice and interpretation. Nevertheless, by introducing a generalized Belinfante-Rosenfeld (BR) energy-momentum tensor we are able to find simple and intuitive arguments to explain why the Minkowski ``canonical'' tensor is more convenient in case (a), whereas for case (b) the Abraham ``kinetic'' tensor turns out to be the best choice. We argue that the Lagrangian model developed here provides a fully consistent framework to understand the long-standing AM controversy in a general and unified way, without unjustified {\it ad hoc} arguments.

The structure of the paper is as follows. Sec.~\ref{lagrangianformalism} summarizes the general field-theoretical framework for the study of open and closed systems, and also presents the definition of the generalized BR energy-momentum tensor. In Sec.~\ref{openelectroo} we turn to the description of the electromagnetic field in a background medium. We show that the Minkowski tensor arises as the BR tensor of such an open system. In Sec.~\ref{BRbalanceelectro} we derive the balance equations for the Minkowski and Abraham tensors and analyse the conditions under which one obtains conservation laws. In Sec.~\ref{dynamicalmediumm} we introduce a Lagrangian model describing the dynamics of the medium as a relativistic fluid with microstructure. Finally, in Sec.~\ref{TBR} we derive the total BR tensor for the closed system of light plus dynamical medium and establish its general decompositions in terms of the Minkowski and Abraham tensors. The last Sec.~\ref{summary} presents the discussion and summary of the results obtained.

Our notation follows \cite{obukhov1} and the book \cite{Birk}. In particular, the indices from the middle of the Latin alphabet $i,j,k,\ldots = 0,1,2,3$ label the 4-dimensional spacetime components, the Latin indices from the beginning of the alphabet $a,b,c,\ldots = 1,2,3$ refer to the 3-dimensional spatial objects and operations (the 3-vectors are also displayed in boldface). The Minkowski metric is defined as $g_{ij}:={\rm diag}(c^2,-1,-1,-1)$. In the 4-dimensional framework spatial components of tensor must be raised or lowered by $g_{ab}=-\delta_{ab}$, but when we are working only with 3-dimensional tensors, we use the convention of using just the Euclidean metric $\delta_{ab}$ to rise and lower spatial indices.

\section{Lagrangian formalism for open and closed systems}\label{lagrangianformalism}

\subsection{Canonical energy-momentum, spin and angular momentum tensors}

We consider a system of $N$ physical fields, which we collectively denote as $\Phi^A(x)$, with $A=1,...,N$ in the four-dimensional flat Minkowski spacetime. Given a Lagrangian density ${\cal L}={\cal L}(\Phi^A(x), \partial_i\Phi^A(x))$, the canonical energy-momentum tensor of the system reads
\begin{equation}
\Sigma{}_{i}{}^{j}:=\frac{\partial {\cal L}}{\partial(\partial_{j} \Phi^A)}\partial_{i}\Phi^A-\delta_{i}^{j}{\cal L}.\label{canonicalemtensor}
\end{equation}
Let us assume that the action of the system $S:=(1/c)\int {\cal L}\ d^4x$ is invariant under spacetime translations, described in terms of local coordinates as $x^i \rightarrow x^i + \varepsilon^i$ with the four infinitesimal parameters $\varepsilon^i$. As a result, the canonical energy-momentum $\Sigma_{i}{}^{j}$ satisfies the Noether identity, which has the meaning of the energy-momentum balance equation:
\begin{equation}
\partial_{j}\Sigma{}_{i}{}^{j} \equiv -\,\frac{\delta{\cal L}}{\delta\Phi^A}\partial_{i}\Phi^A.\label{energymombalance}
\end{equation}
The variational (Euler-Lagrange) derivative is defined as usual by
\begin{equation}\label{EL}
\frac{\delta{\cal L}}{\delta\Phi^A} := \frac{\partial{\cal L}}{\partial\Phi^A} - \partial_i\left(\frac{\partial{\cal L}}{\partial (\partial_i\Phi^A)}\right).
\end{equation}

In addition, let the action $S$ be invariant under infinitesimal Lorentz transformations 
$x^i\rightarrow x^i + \delta x^i$, $\Phi^A\rightarrow \Phi^A + \delta\Phi^A$, with
\begin{equation}\label{lorentz}
\delta x^i = \lambda^i{}_jx^j,\qquad \delta\Phi^A = {\frac 12}\lambda^{kl}(s_{kl})^{A}{}_{B}\Phi^B.
\end{equation}
Here, the infinitesimal parameters $\lambda^{ij}$ are skew-symmetric, $\lambda^{ij} = - \lambda^{ji}$, and  $(s_{kl})^{A}{}_{B}$ are the \emph{Lorentz generators} \cite{footnote} for the fields $\Phi^A$. The Noether theorem then implies the \emph{angular momentum identity}:
\begin{equation}
\partial_{j}S_{kl}{}^{j}-2\Sigma_{[kl]}\equiv-\,\frac{\delta{\cal L}}{\delta \Phi^A}(s_{kl})^{A}{}_{B}\Phi^{B},\label{angularidentity}
\end{equation}
where the \emph{canonical spin current density} is defined by
\begin{equation}
S_{kl}{}^{j}:=\frac{\partial{\cal L}}{\partial(\partial_{j}\Phi^A)}(s_{kl})^{A}{}_{B}\Phi^B.\label{spindef}
\end{equation}
It is convenient to introduce the \emph{canonical orbital angular momentum density} of the system
\begin{equation}
L_{kl}{}^{j}:= x_{k}\,\Sigma_{l}{}^{j} - x_{l}\,\Sigma_{k}{}^{j}.\label{orbitalangularmomdefff}
\end{equation} 
In accordance with the Eq.\,(\ref{energymombalance}), it satisfies the identity 
\begin{equation}
\partial_jL_{kl}{}^j+2\Sigma_{[kl]}\equiv-\,\frac{\delta{\cal L}}{\delta \Phi^A}(l_{kl})^{A}{}_{B}\Phi^{B},\label{orbitalangularidentity}
\end{equation}
with $(l_{kl})^{A}{}_{B}:=\delta^A{}_B\left(x_{k}\,\partial_{l}-x_{l}\,\partial_{k}\right)$ being the usual \emph{orbital angular momentum generators}. Taking the sum of the Eqs.\,(\ref{angularidentity}) and (\ref{orbitalangularidentity}), we obtain the \emph{total angular momentum balance equation}:
\begin{equation}
\partial_{j}J_{kl}{}^{j}\equiv-\,\frac{\delta{\cal L}}{\delta \Phi^A}(j_{kl})^{A}{}_{B}\Phi^B.\label{angularbalance}
\end{equation}
Here we introduced the \emph{canonical total angular momentum density} of the system by
\begin{equation}
J_{kl}{}^{j}:=L_{kl}{}^{j}+S_{kl}{}^{j},\label{Jota}
\end{equation}
and the total angular momentum generators $(j_{kl})^{A}{}_{B}$ for the fields $\Phi^A$ read
\begin{equation}
(j_{kl})^{A}{}_{B}:=(l_{kl})^{A}{}_{B}+(s_{kl})^{A}{}_{B}.\label{totalaop}
\end{equation}
Equations (\ref{energymombalance}) and (\ref{angularbalance}) describe the energy-momentum and angular momentum balance equations, respectively, satisfied by the canonical energy-momentum tensor (\ref{canonicalemtensor}) and the canonical spin tensor (\ref{spindef}). In subsection \ref{openclosed} we will study the conditions under which these two identities (\ref{energymombalance}) and (\ref{angularbalance}) reduce to the {\it conservation laws} of the total energy-momentum and the total angular momentum, respectively.

\subsection{Belinfante-Rosenfeld tensor as a relocalization of the canonical tensor}\label{belinfantetensor}

There is a certain freedom in the definition of the dynamical currents. Namely, it is possible to redefine the energy-momentum and spin tensors without changing the balance (and conservation) laws using the procedure called a {\it relocalization}, see e.g. \cite{hehl76,Schueking}. The original canonical energy-momentum tensor $\Sigma_{i}{}^{j}$ and spin tensor $S_{kl}{}^{j}$ can be replaced by the relocalized ones:
\begin{eqnarray}
\widehat{\Sigma}{}_{i}{}^{j}&:=& \,\Sigma_{i}{}^{j} - \partial_l\,X_{i}{}^{jl},\label{relT}\\
\widehat{S}{}_{kl}{}^{j}&:=& \,S_{kl}{}^{j}-2X_{[kl]}{}^j+\partial_i\,Y_{kl}{}^{ji},\label{relS}
\end{eqnarray}
where the tensors $X_{i}{}^{jl}$ and $Y_{kl}{}^{ji}$ are arbitrary, except for the skew symmetry $X_{i}{}^{jl} = -\,X_{i}{}^{lj}$ and $Y_{kl}{}^{ji} = -\,Y_{kl}{}^{ij} = -\,Y_{lk}{}^{ji}$. One can check straightforwardly that 
\begin{eqnarray}
\partial_{j}\widehat{\Sigma}{}_{i}{}^{j}&\equiv &\partial_{j}\Sigma{}_{i}{}^{j},\label{brcond1}\\
\partial_{j}\widehat{S}{}_{kl}{}^{j}-2\widehat{\Sigma}{}_{[kl]}&\equiv &\, \partial_{j}S_{kl}{}^{j}-2\Sigma_{[kl]},\label{brcond2}
\end{eqnarray}
and accordingly the relocalized energy-momentum and spin satisfy the same balance laws (\ref{energymombalance}), (\ref{angularidentity}) and (\ref{angularbalance}).

A specially useful relocalization is defined by the condition that the relocalized spin $\widehat{S}{}_{kl}{}^{j}$ vanishes. Then the relocalized total angular momentum turns out to be purely orbital $\widehat{J}{}_{kl}{}^{j}=\widehat{L}{}_{kl}{}^{j}:=x_{k}\widehat{\Sigma}_{l}{}^{j} - x_{l}\widehat{\Sigma}_{k}{}^{j}$. From Eq.\,(\ref{relS}), this is achieved when $X_{[kl]}{}^j = (S_{kl}{}^{j} + \partial_i\,Y_{kl}{}^{ij})/2$. Combining this with the two equations obtained by the cyclic permutation of indices, we find
\begin{eqnarray}
X_{i}{}^{jl}&=&-\frac{1}{2}\!\left(S^{jl}{}_{i} + S_{i}{}^{lj} - S_{i}{}^{jl}\right)\nonumber\\
&-&\frac{1}{2}\partial_n\!\left(Y^{jl}{}_i{}^n + Y_{i}{}^{ljn} - Y_{i}{}^{jln}\right).
\end{eqnarray}
This very special relocalization brings us to the definition of the \emph{BR energy-momentum tensor} $\sigma_i{}^j$ \cite{Belinfante1,Belinfante2,Rosenfeld},
\begin{equation}
\sigma_{i}{}^{j}:=\Sigma_{i}{}^{j}+\frac{1}{2}\partial_{k}\left(S^{jk}{}_{i}+S_{i}{}^{kj}-S_{i}{}^{jk}\right).\label{BR}
\end{equation}
Note that $Y_{kl}{}^{ij}$ does not contribute. By construction, (\ref{BR}) satisfies
\begin{equation}
\partial_{j}\sigma_{i}{}^{j}\equiv{}\partial_{j}\Sigma_{i}{}^{j},\qquad
2\sigma_{[ij]}\equiv{} 2\Sigma_{[ij]}-\partial_{k}S_{ij}{}^{k}.\label{beli2}
\end{equation}
Replacing Eq.\,(\ref{beli2}) into Eqs.\,(\ref{energymombalance}), (\ref{angularidentity}) and (\ref{angularbalance}), we see that the canonical spin density $S_{ij}{}^k$ has been absorbed in the new definitions and therefore the BR balance equations turn out to be particularly simple:
\begin{eqnarray}
\partial_{j}\sigma_{i}{}^{j}\equiv{}&-&\frac{\delta{\cal L}}{\delta\Phi^A}\partial_{i}\Phi^A,\label{belienergymom}\\
\partial_{j}l_{kl}{}^{j}\equiv{}&-&\frac{\delta{\cal L}}{\delta \Phi^A}(j_{kl})^{A}{}_{B}\Phi^B,\label{beliamom}\\
2\sigma_{[kl]}&\equiv{}&\frac{\delta{\cal L}}{\delta \Phi^A}(s_{kl})^{A}{}_{B}\Phi^{B}.\label{beliarelation}
\end{eqnarray}
By construction, the \emph{total BR angular momentum density} 
\begin{equation}
l_{kl}{}^{j} := x_{k}\sigma_{l}{}^{j} - x_{l}\sigma_{k}{}^{j},\label{amombelinfantee}
\end{equation}
is purely orbital. Interestingly, the spin terms contained in the BR tensor (\ref{BR}) produce observable torques, as it has been recently predicted for the interaction of evanescent optical fields with a spinning Mie particle \cite{Bliokh2014}.

An additional very remarkable feature of the BR energy-momentum tensor (\ref{BR}) is that it is invariant under \emph{canonical transformations}, i.e. under a shift of the Lagrangian by a total derivative of the form ${\cal L} \rightarrow {\cal L} + \partial_i\xi^i$, with $\xi=\xi(\Phi^A(x))$. From the definition  (\ref{BR}) it is straightforward to show that the BR tensor corresponding to the Lagrangian $\overline{{\cal L}}:=\partial_i\xi^i$ vanishes identically, whereas the canonical one (\ref{canonicalemtensor}) is nonzero. As a result, the BR tensors constructed from ${\cal L}$ and ${\cal L}+\partial_i\xi^i$ are the same, whereas the canonical energy-momentum (\ref{canonicalemtensor}) and spin (\ref{spindef}) tensors are different in the two cases. In this sense, and in the same way as the equations of motion, the BR tensor is uniquely defined for a given dynamical system.

In contrast to the standard literature \cite{landau,jackson} where the BR tensor is introduced mainly as a symmetrization procedure for a canonical energy-momentum tensor in vacuum, here we show that the BR tensor can always be introduced, independently if the system is open or closed. This is a key point that allows us to better understand the structure underlying the electrodynamics in moving material media and the Abraham-Minkowski controversy in particular.   

\subsection{Open and closed systems}\label{openclosed}

We say that a system is \textit{closed} if it is not influenced by physical fields from the outside of the system. In a Lagrangian field theory, a closed system is the one in which the dynamics of all the fields is completely determined by the fields of the system through the Euler-Lagrange equations derived from the action principle (i.e. it is a ``self-interacting" system, not coupled to the external world). The fields that satisfy Euler-Lagrange equations $\delta{\cal L}/\delta\Phi^A_{\rm dyn}=0$ are called \emph{dynamical fields} $\Phi^A_{\rm dyn}$ and therefore a closed system is one where all its fields are dynamical. On the other hand, a system is said to be {\it open}, if its dynamics is not only determined by dynamical fields, but also by \emph{external fields}, which are given functions of space and time whose values are not affected by the evolution of the system. The external fields do not satisfy Euler-Lagrange equations and therefore they can also be understood as non-dynamical or ``background'' fields.

It is important to notice that in the discussion above, the set $\Phi^A$ includes in general both dynamical and external fields. It is then convenient to explicitly distinguish between them by splitting the set of all matter fields into two subsets: $\Phi^A=\left\lbrace\Phi^{\cal A}_{\rm dyn},\Phi^\alpha_{\rm ext}\right\rbrace$, with ${\cal A}=1,...,N_{\rm dyn}$ and $\alpha=1,...,N_{\rm ext}$, such that $N_{\rm dyn}+N_{\rm ext}=N$. Considering this separation of the fields in the balance equations (\ref{belienergymom})-(\ref{beliarelation}) and using the fact that $\Phi_{\rm dyn}^{\cal A}$ are solutions of the Euler-Lagrange equations, $\delta {\cal L}/\delta \Phi^{\cal A}_{\rm dyn}=0$, $\forall {\cal A}$, the balance equations reduce to
\begin{eqnarray}
\partial_{j}\sigma_{i}{}^j&={}&-\frac{\delta{\cal L}}{\delta\Phi^\alpha_{\rm ext}}\partial_{i}\Phi^\alpha_{\rm ext},\label{belienergymom3}\\
\partial_{j}l_{kl}{}^{j}&={}&-\frac{\delta{\cal L}}{\delta\Phi^\alpha_{\rm ext}}(j_{kl})^{\alpha}{}_{\beta}\Phi^\beta_{\rm ext},\label{beliamom3}\\
2\sigma_{[kl]}&={}&\frac{\delta{\cal L}}{\delta\Phi^\alpha_{\rm ext}}(s_{kl})^{\alpha}{}_{\beta}\Phi^{\beta}_{\rm ext}.\label{beliarelation3}
\end{eqnarray}
In Eqs.\,(\ref{belienergymom3})-(\ref{beliarelation3}) we also assumed that $(s_{kl})^{\alpha}{}_{\cal A}=0$, since the separation of fields into dynamical and external ones must be preserved under Lorentz transformations. Note also that the metric must be excluded from the dynamical/external field discussion, since it is a field invariant under Lorentz and translational transformations; and therefore it has no contribution on the right hand side of the balance equations (\ref{belienergymom3})-(\ref{beliarelation3}).

\subsubsection{Closed system}\label{closedddd}

For a closed system, the Lagrangian density ${\cal L}$ does not depend on external fields and hence the right hand sides of the BR balance equations (\ref{belienergymom})-(\ref{beliarelation}) vanish. As a consequence, the BR energy-momentum tensor $\sigma_{i}{}^{j}$ and the BR orbital angular momentum density $l_{kl}{}^{j}$ are always \emph{conserved} in closed systems:
\begin{equation}
\partial_{j}\sigma_{i}{}^{j}={}0,\qquad
\partial_{j}l_{kl}{}^{j}={}0,\label{conservationclose}
\end{equation}
and, in addition, $\sigma_{i}{}^{j}$ is \emph{symmetric}:
\begin{equation}
\sigma_{[kl]}={}0.\label{BRsymmetricclosed}
\end{equation} 

In a closed system, also the right hand side of the canonical balance equations (\ref{energymombalance}), (\ref{angularidentity}) and (\ref{angularbalance}) vanish, and therefore the canonical energy-momentum tensor $\Sigma_{j}{}^{i}$, together with the total canonical angular momentum density $J_{kl}{}^{j}=S_{kl}{}^{j}+L_{kl}{}^{j}$ are conserved as well:
\begin{equation}
\partial_{j}\Sigma_{i}{}^{j}={}0,\qquad \partial_{j}J_{kl}{}^{j}=0.\label{conservedclosecanonical}
\end{equation}
However, and in contrast to the BR tensor, the canonical tensor of a closed system is \emph{not symmetric, in general}:
\begin{equation}
2\Sigma_{[kl]}={}\partial_{j}S_{kl}{}^{j}\neq 0.\label{nonsymmetrysigma}
\end{equation} 
The asymmetry of the canonical energy-momentum tensor of a closed system is a consequence of the tensorial nature of the dynamical fields. This is completely consistent in the Lagrangian formalism with the conservation of energy-momentum and total angular momentum (\ref{conservedclosecanonical}). In the literature one can sometimes find a `proof' that an energy-momentum tensor must be symmetric in closed systems in order to be consistent with the conservation of angular momentum, see \cite{MTW,kundu}, for example. However, in such a proof one tacitly assumes that the system does not have microstructure, in other words that all the dynamical fields are scalars. In the latter case the canonical tensor of a closed system must indeed be symmetric, since the spin density in Eq.\,(\ref{nonsymmetrysigma}) vanishes. This is well known in classical field theory; see for instance Refs.\,\cite{hehl76,Schueking,TT}.

\subsubsection{Open system}\label{opensystemmm}

When the system is open, we have to keep the corresponding terms on the right hand sides of the balance equations and therefore the canonical energy-momentum tensor $\Sigma_{i}{}^{j}$ and the BR tensor $\sigma_{i}{}^{j}$ are \emph{neither symmetric  nor conserved, in general}. The non-vanishing terms in the balance equations are proportional to the variational derivatives of $\cal L$ with respect to the external or non-dynamical fields, present in the open system. They describe forces and torques which result from the interaction of the system with  the external fields and hence the asymmetry and non-conservation of the energy-momentum tensors in open systems is \textit{necessary} for the correct and consistent description of these systems within the Lagrange-Noether formalism. For a coordinate-free analysis of this issue see \cite{ann2012}. 

It is important to mention that \emph{conserved quantities} like energy, momentum and angular momentum do exist for certain classes of open systems. These conserved quantities are related to the \emph{symmetries} presented by the specific external fields $\Phi^A_{\rm ext}$ that act on the open system. Let us now recall the physical interpretation of the components of the BR energy-momentum tensor $\sigma_i{}^j$. Following \cite{obukhov1}, we make a $(1+3)$ decomposition:
\begin{equation}
\sigma_{i}{}^{j}:=\left(\begin{array}{cc}{\cal U} & S^a\\ 
-\pi_a & -p_a{}^b\end{array}\right),\label{identificationem}
\end{equation}
where ${\cal U}$ is the energy density, $S^a$ the energy flux density, $\pi_a$ the momentum density and $p_a{}^b$ the momentum flux density of the open system. The inspection of Eqs.\,(\ref{belienergymom3})-(\ref{beliarelation3}) shows that under certain symmetry conditions conserved quantities do exist. In particular, suppose that all the external fields are time-independent, i.e. that  $\partial \Phi^\alpha_{\rm ext}/\partial t=0$, for all values of $\alpha$. Then from Eq.\,(\ref{belienergymom3}) with $i=0$, and using Eq.\,(\ref{identificationem}), we find an energy continuity equation:
\begin{equation}
\frac{\partial {\cal U}}{\partial t}+\partial_{a}S^a=0.\label{dUdS}
\end{equation}
Similarly, if all the external fields are invariant under spatial translations in the $x^a$ direction, i.e. $\partial_a\Phi^\alpha_{\rm ext}=0$, for all values of $\alpha$, then Eq.\,(\ref{belienergymom3}) with $i=a$ yields the momentum continuity equation of the open system,
\begin{equation}
\frac{\partial \pi_a}{\partial t}+\partial_{b}p_a{}^b=0.\label{dpdp}
\end{equation}
The Lorentz transformations around spatial axes correspond just to spatial rotations. Suppose that all external fields are invariant under spatial rotations in some plane $[a,b]$, with $a,b=1,2,3$, such that $(j_{ab})^\alpha{}_\beta\Phi^\beta_{\rm ext}=0$, for all values of $\alpha$. Then we find the orbital angular momentum continuity equation in the plane $[a,b]$:
\begin{equation}
\frac{\partial l_{ab}{}^{0}}{\partial t}+\partial_cl_{ab}{}^c=0.\label{angularmomcont7}
\end{equation}
In Sec.\,\ref{asymmetrymink} we will see that $l_{ab}{}^{0}$ and $l_{ab}{}^c$ in Eq.\,(\ref{angularmomcont7}) are related to the vector angular momentum density $\bm{l}:=\bm{x}\times\bm{\pi}$ and the angular momentum flux density $K_a{}^b$ of the open system, respectively. Integrating the continuity equations (\ref{dUdS}), (\ref{dpdp}), and (\ref{angularmomcont7}) over a 3-spatial region $V$, we obtain that the rate of temporal change of the total energy, momentum, and angular momentum inside $V$ is compensated by the corresponding fluxes through the boundary $\partial V$. Therefore, when there is no flux through $\partial V$, conserved quantities are found (constants of motion).

Finally, let us assume that all the external fields are invariant under boosts described by the  Lorentz transformations in the time-space plane corresponding to $\lambda^{0a}$, such that $(j_{0a})^\alpha{}_\beta\Phi^\beta_{\rm ext}=0$, for all values of $\alpha$. Then we obtain the continuity equation
\begin{equation}
\frac{\partial l_{0a}{}^{0}}{\partial t}+\partial_bl_{0a}{}^b=0.\label{centerofenergy7}
\end{equation}
It is more difficult to find a physical interpretation of Eq.\,(\ref{centerofenergy7}), since it is not easy to find an example of an external field that is invariant under boosts (besides the ones constructed only with the Minkowski metric). In this work, i.e. the case of the electromagnetic field in material media, we will see in Sec.~\ref{asymmetrymink} that the very presence of the medium breaks the symmetry under boosts and therefore $l_{0a}$ are never conserved quantities in that open system. The exception is of course the case of light in vacuum, when the system is actually a closed one.  

With regard to the vanishing of the antisymmetric part of the BR tensor in open systems $\sigma_{[ij]}$, we clearly see from Eq.\,(\ref{beliarelation3}) that it is also determined by properties of the external fields $\Phi^{\alpha}_{\rm ext}$. In particular, in a system where all the external fields are scalars, we have $(s_{kl})^\alpha{}_\beta=0$, and thus Eq.\,(\ref{beliarelation3}) trivially yields $\sigma_{[kl]}=0$. Nevertheless, it is important to stress here, the symmetry of the full energy-momentum 4-tensor is not a restriction for the conservation of the angular momentum, as it is sometimes stated in the literature. In fact, only the antisymmetric part of the spatial components $\sigma_{[ab]}$ plays a role in this, as can be inferred from Eqs.\,(\ref{totalaop}), (\ref{beliamom3}) and (\ref{beliarelation3}). For example, in Sec.\,\ref{asymmetrymink} we discuss the case of light propagating inside an homogeneous and isotropic medium at rest, for which the BR 4-tensor is not symmetric, but its spatial components form a 3-dimensional symmetric tensor and the corresponding BR orbital angular momentum is a conserved quantity. 

\section{Electromagnetic field in matter as an open system}\label{openelectroo}

We begin our discussion by considering the electromagnetic field in matter as an \emph{open} system, in which only the electromagnetic field is assumed to have dynamics described by the \emph{macroscopic} Maxwell equations. This is the case if the influence of the electromagnetic field on the macroscopic dynamics of the medium is negligible or if an external agent keeps the medium in a predetermined state of motion, independently of the values of the electromagnetic field. The validity of this approach is the same as the usual macroscopic electromagnetic theory \cite{Birk,jackson,LL} with the continuum hypothesis assumed; no atomic systems will be studied in this framework. 

The continuous material medium is considered as a \emph{fixed} ``background'', whose optical and electromagnetic properties are specified by non-dynamical constitutive relations. The latter are phenomenological equations that describe the macroscopic response of a material medium under the action of the electromagnetic field. Mathematically, they can be specified by relating the electric and magnetic excitations $\bm{D}$ and $\bm{H}$ to the electric and magnetic fields $\bm{E}$ and $\bm{B}$:
\begin{equation}
\bm{D}={}\bm{D}[\bm{E},\bm{B}],\qquad
\bm{H}={}\bm{H}[\bm{E},\bm{B}].\label{generalrelationss}
\end{equation}
These relations are determined by the electromagnetic properties of the medium.  All the assumptions regarding the induced electric and magnetic dipole moments within the medium are already taken into account in the constitutive relations  (\ref{generalrelationss}). After solving the problem, these quantities can be calculated from
\begin{equation}
\bm{P}={\bm D}-\varepsilon_0{\bm E},\quad
\bm{M}=\bm{B}/\mu_0-\bm{H}\label{PandM},
\end{equation}
where $\bm{P}$ is the polarization and $\bm{M}$ the magnetization of the medium, induced by the electromagnetic field.

\subsection{Covariant constitutive relations for linear and non-dispersive media}\label{covariantconsti}

Throughout this work we consider the case of media with constitutive laws (\ref{generalrelationss}) that are linear in the fields and local in space and time. Therefore, we do not include light dispersion effects \cite{Post,Furutsu}, but we will be able to describe in full generality anisotropy and birefringence effects \cite{Post,ORR02,Lindell,itin,favaro,dahl}, magneto-electric effects \cite{Odell,obukhovhehl2007}, inhomogeneities, dissipation, and effects due to the motion of the medium \cite{dielectricslab,penfieldhaus1967,Birk}. The most general constitutive relations for a linear and non-dispersive medium read
\begin{eqnarray}
D^a&={}&\varepsilon_0\varepsilon^{ab}E_b+\beta^{a}{}_{b}B^b,\label{explicitconst1}\\
H_a&={}&\alpha_{a}{}^{b}E_b+\mu_0^{-1}(\mu^{-1})_{ab}B^b.\label{explicitconst2}
\end{eqnarray}
Here $\varepsilon^{ab}$ is the relative permittivity or dielectric 3-tensor of the medium, $(\mu^{-1})_{ab}$ is the inverse relative permeability tensor and $\alpha_{a}{}^{b}$, $\beta^{a}{}_{b}$ are the linear magneto-electric coupling coefficients. The quantities $\varepsilon^{ab}$ and $(\mu^{-1})_{ab}$ are dimensionless, whereas $\alpha_{a}{}^{b}$ and $\beta^{a}{}_{b}$ have dimensions of $\sqrt{\varepsilon_0/\mu_0}$, where $\varepsilon_0$ and $\mu_0$ are the permittivity and permeability of vacuum, respectively. Each of the four matrices have $9$ components in general, so the most general linear and non-dispersive medium is characterized by $36$ real functions. To describe dispersion effects, one needs to consider a non-local in time and/or space response of the medium, and as a result the constitutive relations in Eqs.\,(\ref{explicitconst1})-(\ref{explicitconst2}) would have to be replaced by convolution integrals \cite{Post,Furutsu}. The dispersive constitutive relations become local in terms of the Fourier transformed electromagnetic and material variables, and one can generalize the present framework accordingly. For more details on a possible generalization, see Ref.\,\cite{Furutsu}.

The general constitutive relations (\ref{explicitconst1})-(\ref{explicitconst2}) can be expressed in covariant form \cite{obukhov1,ORR02} as
\begin{equation}
H^{ij} = {\frac 12}\chi^{ijkl}F_{kl}.\label{4const}
\end{equation}
Here $H^{ij}$ is the electromagnetic excitation 4-tensor and $F_{ij}$ is the electromagnetic field strength, whose components are identified as in \cite{obukhov1} by
\begin{eqnarray}
F_{a0}&:={}&E_a,\qquad F_{ab}:=\epsilon_{abc}B^{c},\\
H^{a0}&:={}&-D^a,\qquad H^{ab}:=\epsilon^{abc}H_{c}.
\end{eqnarray}
The external field $\chi^{ijkl}$ is the so-called \emph{constitutive tensor} with $36$ 
independent components which encode the real constitutive functions of the medium. 
Historically, this general constitutive relation was first formulated
by Bateman \cite{Bateman}, Tamm \cite{Tamm1,Tamm2,Tamm3}, and later in the 
modern notation by Post \cite{Post}. 
Since $H^{ij}$ and $F_{ij}$ in Eq.\,(\ref{4const}) are both antisymmetric, $\chi^{ijkl}$ must be, by definition, also antisymmetric in the two first and the two last indices, i.e.,
\begin{equation}
\chi^{ijkl}={}-\chi^{jikl},\qquad
\chi^{ijkl}={}-\chi^{ijlk},\label{symmetry2}
\end{equation}
thus indeed $\chi^{ijkl}$ has $36$ independent components. If we decompose Eq.\,(\ref{4const}) in temporal and spatial components and compare it with Eqs.\,(\ref{explicitconst1})-(\ref{explicitconst2}), we can identify $\chi^{ijkl}$ in a more familiar way with the 3-matrices:
\begin{eqnarray}
\chi^{0ab0}&={}&\varepsilon^{ab}/\mu_0,\label{invidenti1}\\
\chi^{abcd}&={}&\epsilon^{abe}\epsilon^{cdf}(\mu^{-1})_{ef}/\mu_0,\label{invidenti2}\\
\chi^{bc0a}&={}&-c\ \epsilon^{dbc}\alpha_d{}^a,\\
\chi^{0abc}&={}&c\ \epsilon^{dbc}\beta^a{}_d.\label{invidenti4}
\end{eqnarray}
In fact, there is a one-on-one correspondence between $\chi^{ijkl}$ and $\varepsilon^{ab}$, $(\mu^{-1})_{ab}$, $\alpha_{a}{}^{b}$ and $\beta^{a}{}_{b}$, since one can easily invert the relations (\ref{invidenti1})-(\ref{invidenti4}).

In addition, when the medium is non-dissipative (lossless), the {\it real} constitutive tensor must satisfy an extra symmetry
\begin{equation}
\chi^{ijkl}=\chi^{klij},\label{extrasym}
\end{equation}
which reduces the number of its independent components to $21$. This additional condition (\ref{extrasym}) implies that $\varepsilon^{ab}$ and $(\mu^{-1})_{ab}$ must be symmetric in a non-dissipative medium, whereas $\alpha_{a}{}^{b}$ must be the negative transpose of $\beta^{a}{}_{b}$, i.e. $\alpha_{a}{}^{b}=-\beta^{b}{}_{a}$. Examples of constitutive tensors can be found in \cite{dielectricslab,obukhov1} for an isotropic moving medium, also in \cite{obukhovhehl2007} for a magneto-electric medium and recently in \cite{ORR02} for a liquid crystal as a specific uniaxial anisotropic dielectric and diamagnetic medium. Another simple, but physically important case is the constitutive tensor of the vacuum, which reads
\begin{equation}
\chi^{ijkl}_{(\rm vac)}:=(g^{ik}g^{jl}-g^{il}g^{jk})/\mu_0,\label{constvacuum2}
\end{equation}
where the inverse Minkowski spacetime metric is given by $g^{ij}={\rm diag}(1/c^2,-1,-1,-1)$. Substituting Eq.\,(\ref{constvacuum2}) into Eq.\,(\ref{4const}), we obtain the well-known vacuum relations $\bm{D}=\varepsilon_0\bm{E}$ and $\bm{H}=\bm{B}/\mu_0$.

\subsection{Lagrangian formalism for the electromagnetic field in a non-dynamical background medium}\label{canonicalemsection}

The Lagrangian density for the electromagnetic field in matter has the covariant form \cite{obukhov1,ORR02}
\begin{equation}
{\cal L}^{\rm e}:={}-\frac{1}{4}F_{ij}H^{ij}=-\,\frac{1}{8}\chi^{ijkl}F_{ij}F_{kl}.\label{densidadlag22}
\end{equation}
If one defines the electromagnetic 4-potential $A_{i}$, such that $F_{ij}:= \partial_iA_j-\partial_jA_i$, then the homogeneous Maxwell equations are automatically satisfied. The inhomogeneous Maxwell equations without sources are obtained as Euler-Lagrange equations for $A_i$:
\begin{equation}\label{Maxwellsequations}
\frac{\delta{\cal L}^{\rm e}}{\delta A_{i}}=-\partial_{j}H^{ij}=0.
\end{equation}
It is important that only \emph{non-dissipative} media can be analyzed with this Lagrangian formalism, due to the condition (\ref{extrasym}) fulfilled by the $\chi^{ijkl}$ in Eq.\,(\ref{densidadlag22}).

Using the general definition (\ref{canonicalemtensor}), we compute the canonical energy-momentum tensor of the electromagnetic field in matter from the Lagrangian (\ref{densidadlag22}), obtaining
\begin{equation}\label{canonicalem}
{\stackrel{\rm e} \Sigma}_{i}{}^{j}={}{\Theta}_{i}{}^{j}+H^{kj}(\partial_{k}A_{i}).
\end{equation}
Here, ${\Theta}_{i}{}^{j}$ is the well-known \textit{Minkowski energy-momentum tensor}, defined in any medium as
\begin{equation}
{\Theta}_{i}{}^{j}:=F_{ik}H^{kj}+{\frac 14}\delta^{j}_{i}F_{kl}H^{kl}.\label{minkowskitensor}
\end{equation}
Notice that the electromagnetic canonical tensor ${\stackrel{\rm e}\Sigma}_{i}{}^{j}$ is gauge non-invariant due to the last term.

\subsection{Minkowski tensor as the Belinfante-Rosenfeld tensor of light in matter}\label{subminkbeli}

In order to construct the BR tensor of the electromagnetic field in matter, we first calculate the spin density using the general definition (\ref{spindef}) and the Lagrangian (\ref{densidadlag22}), obtaining ${\stackrel {\rm e}S}_{kl}{}^{j}=H^{j}{}_{l}A_k - H^{j}{}_{k}A_{l}$. Then, using ${\stackrel {\rm e}S}_{kl}{}^{j}$, the canonical tensor (\ref{canonicalem}) and the equations of motion (\ref{Maxwellsequations}) in the definition (\ref{BR}), we see that the BR tensor of the electromagnetic field in matter ${\stackrel {\rm e}\sigma}_{i}{}^{j}$ actually coincides with the Minkowski one:
\begin{equation}
{\stackrel {\rm e}\sigma}_{i}{}^{j}={\Theta}_{i}{}^{j}.\label{brminkowski}
\end{equation}
By construction, the BR and canonical tensors satisfy the same balance equations. Nevertheless, the important advantage of ${\stackrel {\rm e}\sigma}_{i}{}^{j}$ as compared to ${\stackrel{\rm e} \Sigma}_{i}{}^{j}$ is that the gauge non-invariant term $H^{kj}\partial_{k}A_{i}$ in Eq.\,(\ref{canonicalem}) has now disappeared. In Sec.~\ref{belinfantetensor}, we also demonstrated that the BR tensor is invariant under the redefinition of  the electromagnetic Lagrangian (\ref{densidadlag22}) by adding arbitrary total derivative terms, in contrast to the canonical tensor.

One can also take into account the sources of the electromagnetic field in the Lagrangian formalism by adding the Lagrangian ${\cal L}^{\rm J}:=J^{i}A_{i}$ to the electromagnetic one in Eq.\,(\ref{densidadlag22}). Here $J^{i}:=(\rho,\bm{j})$ is the \emph{external} 4-current density, with $\rho$ being the external charge density and $\bm{j}$ the external current density. One can easily check that the inhomogeneous Maxwell equations, $\partial_jH^{ij}=J^{i}$, are obtained as Euler-Lagrange equations for $A_i$,  when considering the electromagnetic Lagrangian with sources ${\cal L}^{'\rm e}:={\cal L}^{\rm e}+{\cal L}^{\rm J}$. Due to the additional source term, the BR tensor corresponding to ${\cal L}^{'\rm e}$  is no longer gauge invariant and also does not coincide with the Minkowski tensor $\Theta_i{}^j$:
\begin{equation}
{\stackrel {\rm e}\sigma}{}'_{i}{}^{j}={\Theta}_{i}{}^{j}-\delta^{j}{}_{i}J^{k}A_{k}+A_{i}J^{j}.\label{Belinfantemink2}
\end{equation}
Nevertheless, as we will show in the next section, even when $J^i\neq 0$ all the gauge non-invariant terms go away in the electromagnetic energy-momentum balance equation.

\section{Electromagnetic balance equations and conserved Minkowski quantities}\label{BRbalanceelectro}

\subsection{Energy and momentum of the electromagnetic field in a background medium}\label{balanceenergymom}

Given the explicit form of the BR tensors in Eqs.\,(\ref{brminkowski})-(\ref{Belinfantemink2}), we are now in a position to analyze the general BR balance equation (\ref{belienergymom3}) in the case of the electromagnetic field in background matter. After evaluating its right hand side, which depends on the external fields acting on the system, we obtain
\begin{equation}
\partial_{j}{\stackrel {\rm e}\sigma}{}'_{i}{}^{j} ={\frac 18}F_{mj}F_{kl}\partial_{i}\chi^{mjkl} - A_j\partial_iJ^j,
\end{equation}
and rearranging the terms, we find that all the gauge non-invariant terms are gone
\begin{equation}
\partial_{j}{\Theta}_{i}{}^{j} + {\cal F}_i^{\rm J} + {\cal F}_i^{\rm m} = 0.\label{divthetaonshell}
\end{equation}
Here,
\begin{equation}
{\cal F}_i^{\rm J}:=-F_{ij}J^{j},\label{4lorentzforce}
\end{equation}
is the usual Lorentz 4-force density, which describes, in a covariant way, the rate of energy and momentum transfer from the electromagnetic field to the external charges and currents inside an infinitesimal volume element. The components of ${\cal F}_i^{\rm J}$ can be explicitly identified as ${\cal F}_i^{\rm J}:=(w^{\rm J},-\bm{f}^{\rm J})$, where $w^{\rm J}:=\bm{j}\cdot\bm{E}$ is the electromagnetic work on the external currents and $\bm{f}^{\rm J}:=\rho\bm{E}+\bm{j}\times\bm{B}$ is the Lorentz force on external charges and currents. In accordance with Eq.\,(\ref{divthetaonshell}), the conversion from electromagnetic into mechanical energy and momentum is balanced by a corresponding rate of decrease of energy and momentum of the electromagnetic field in matter, which is quantified by the 4-divergence of the Minkowski energy-momentum tensor $\Theta_i{}^j$.

The last term in Eq.\,(\ref{divthetaonshell}) reads explicitly 
\begin{equation}
{\cal F}_i^{\rm m}:={}-{\frac 18}F_{mj}F_{kl}\partial_{i}\chi^{mjkl},\label{effforce1}
\end{equation}
which is quadratic in the electromagnetic field and linear in the derivatives of the constitutive material tensor $\chi^{ijkl}(\bm{x},t)$. Suppose now that the medium is homogeneous in space ($\partial_{a}\chi^{ijkl}=0$) and time-independent ($\partial_t \chi^{ijkl}=0$), such that ${\cal F}_i^{\rm m}$ vanishes. Then, if additionally the external charges and currents are absent, $J^i=0$, the Minkowski tensor $\Theta_i{}^j$ is conserved: its components satisfy continuity equations, $\partial_j\Theta_i{}^j=0$. In other words, for vanishing free charges and currents, the conditions for the conservation of the Minkowski tensor (\ref{minkowskitensor}) of light in matter are \emph{directly determined by the space and time translational symmetry of the background material medium}. The latter is not surprising since when $J^i=0$, the only remaining external field is $\chi^{ijkl}$, so its symmetries will determine the conserved quantities in this open system.

It is convenient to interpret the term ${\cal F}_i^{\rm m}$ in the balance equation (\ref{divthetaonshell}) also as a 4-force density, ${\cal F}_i^{\rm m}=(w^{\rm m},-\bm{f}^{\rm m})$, which describes the macroscopic transfer of energy and momentum from the electromagnetic field to the \textit{bound charges and currents} of the background material medium. We may call ${\cal F}_i^{\rm m}$ the \textit{effective material 4-force density}, since the light propagating in an homogeneous region of the medium would not exert any macroscopic ``effective'' force on it. This is completely consistent with the very basic and well-known observation that in a linear, non-dissipative, non-dispersive, homogeneous and time-independent medium at rest, with no external charges and currents, a light pulse propagates with constant amplitude and velocity $\bm{v}=c\bm{\hat{k}}/n$. Of course there are microscopic interactions between the electromagnetic field and the atoms/molecules of the medium, but macroscopically the light in this kind of medium moves as if there were no effective force. On the contrary, if light propagates on an inhomogeneous medium, for instance when light encounters an interface between two different homogeneous media, then the electromagnetic wave would exert an effective macroscopic force on the medium and therefore change its velocity of propagation (refraction and reflection at the boundary). As a result, the action of this effective force only at the boundary is consistent with the change of Minkowski momentum there and also with its conservation when there is no force in a locally homogeneous medium.  

It is worthwhile to note that the energy-momentum balance equation (\ref{divthetaonshell}) can also be derived without the Lagrangian formalism, if we just use the Maxwell equations in the definition of the Lorentz 4-force density (\ref{4lorentzforce}) on the external charges and currents. In the trivial, but very important case when the medium is the vacuum (\ref{constvacuum2}), the energy-momentum balance equation (\ref{divthetaonshell}) reduces to the standard one, $\partial_j{\stackrel {\rm vac}\Theta}_i{}^j=F_{ij}J^{j}$ \cite{jackson}, with
\begin{equation}
{\stackrel {\rm vac}\Theta}_i{}^j:=\left(F_{ik}F^{kj}+(1/4)\delta_i{}^jF_{kl}F^{kl}\right)/\mu_0,\label{symmetricminkowski}
\end{equation}
being the well-known symmetric Minkowski tensor in vacuum. As a consequence, when $J^i=0$, the Minkowski tensor in vacuum will be always conserved.

\subsubsection{Explicit Minkowski energy balance equation}

Using the identifications of the components of a energy-momentum tensor (\ref{identificationem}), we can decompose Eq.\,(\ref{divthetaonshell}) in space and time components and find separate balance equations for the electromagnetic energy and momentum. They are valid for any linear, non-dissipative and non-dispersive background medium (not necessarily static). Evaluating Eq.\,(\ref{divthetaonshell}) for $i=0$ and using the identifications (\ref{identificationem}), we obtain
\begin{equation}
\frac{\partial {\cal U}^{\rm M}}{\partial t}+\bm{\nabla}\cdot\bm{S}^{\rm M}+w^{\rm J}+w^{\rm m}=0.\label{energybal}
\end{equation}
Here the Minkowski energy density ${\cal U}^{\rm M}$, the energy flux density $\bm{S}^{\rm M}$, the power density transferred to the external currents $w^{\rm J}$ and the material effective power density transferred to the bound charges and currents $w^{\rm m}$, are all explicitly given by 
\begin{eqnarray}
{\cal U}^{\rm M}&={}&\frac{1}{2}\left(\bm{E}\cdot\bm{D}+\bm{H}\cdot\bm{B}\right),\\
\bm{S}^{\rm M}&={}&\bm{E}\times\bm{H},\\
w^{\rm J}&={}&\bm{j}\cdot\bm{E},\label{wJ}\\
w^{\rm m}&={}&\frac{1}{2}\varepsilon_0E_aE_b\frac{\partial}{\partial t}\varepsilon^{ab}-\frac{1}{2\mu_0}B^aB^b\frac{\partial}{\partial t}(\mu^{-1})_{ab}\nonumber\\
&&+E_aB^b\frac{\partial}{\partial t}\beta^a{}_b.\label{wm}
\end{eqnarray}
Consider now a spatial region $V$ bounded by a closed surface $\partial V$ within the medium with time-independent properties and free of external currents, $\bm{j}=0$. We then obtain an integral energy conservation equation
\begin{equation}
\frac{d}{dt}\int_V d^3x\ {\cal U}^{\rm M}=-\oint_{\partial V} dS\ \left(\bm{\hat{n}}\cdot\bm{S}^{\rm M}\right),
\end{equation} 
where $\bm{\hat{n}}$ is the unit vector field normal to the boundary $\partial V$. When the electromagnetic energy flux $\bm{S}^{\rm M}$ through the boundary of the volume $V$ vanishes, the integrated Minkowski energy inside this region is a time-independent 
quantity:
\begin{equation}
E^{\rm M}:=\frac{1}{2}\int_V d^3x\ \left(\bm{E}\cdot\bm{D}+\bm{H}\cdot\bm{B}\right)={\rm constant}.\label{conservedener5}
\end{equation}

\subsubsection{Explicit Minkowski momentum balance equation}

On the other hand, evaluating Eq.\,(\ref{divthetaonshell}) for the spatial components $i=a$, we get the momentum balance equation for the electromagnetic field:
\begin{equation}\label{balmom}
\frac{\partial\pi_a^{\rm M}}{\partial t}+\partial_b{\stackrel {\rm M}p}_a{}^b + f_a^{\rm J}+f^{\rm m}_a=0.
\end{equation}
Here $\bm{\pi}^{\rm M}$ is the Minkowski momentum density of the electromagnetic field, ${\stackrel {\rm M}p}_a{}^b$ the Minkowski electromagnetic stress tensor, $\bm{f}^{\rm J}$ is the Lorentz force density exerted by the field on the external charges and currents and $\bm{f}^{\rm m}$ is the effective material force density exerted by the field on the bound charges and currents of the medium:
\begin{eqnarray}
\bm{\pi}^{\rm M}&={}&\bm{D}\times\bm{B},\label{momentM}\\ \label{stressM}
{\stackrel {\rm M}p}_{a}{}^{b}&={}&-E_aD^b-H_aB^b+\frac{1}{2}\delta^b_a\left(E_cD^c+H_cB^c\right),\\
\bm{f}^{\rm J}&={}&\rho\bm{E}+\bm{j}\times\bm{B},\\
f^{\rm m}_a&={}&-\frac{1}{2}\varepsilon_0E_bE_c\ \partial_{a}\varepsilon^{bc}+\frac{1}{2\mu_0}B^bB^c\ \partial_{a}(\mu^{-1})_{bc}\nonumber\\
&&-E_bB^c\ \partial_{a}\beta^b{}_c.\label{mforce}
\end{eqnarray}
Specializing again to a spatial region $V$ within an homogeneous medium without external charges and currents ($\rho=0$, $\bm{j}=0$), with all the derivatives in Eq.\,(\ref{mforce}) vanishing, we find an integral momentum conservation equation:
\begin{equation}
\frac{d}{dt}\int_V d^3x\ \pi_a^{\rm M}=-\oint_{\partial V} dS\ {\stackrel {\rm M}p}_a{}^b\hat{n}_b.
\end{equation} 
In the absence of electromagnetic momentum flux through the boundary $\partial V$, the integrated Minkowski momentum is a time-independent quantity:
\begin{equation}
\bm{p}^{\rm M}:=\int_V d^3x\ \bm{D}\times\bm{B}={\rm constant}.\label{conservedmom5}
\end{equation}

The two constants of motion of this open system (under translational symmetry conditions of the medium in space and time) turn out to be the Minkowski energy (\ref{conservedener5}) and the Minkowski momentum (\ref{conservedmom5}). They both are \emph{effective} quantities that depend not only on the electric and magnetic fields $\bm{E}$ and $\bm{B}$, but also on the medium properties $\varepsilon^{ab}$, $(\mu^{-1})_{ab}$ and $\beta^a{}_b$ through the electric and magnetic excitations $\bm{D}$ and $\bm{H}$.

\subsection{Angular momentum  and asymmetry of the Minkowski tensor}\label{asymmetrymink}

Since the Minkowski tensor $\Theta_i{}^j$ is the BR energy-momentum tensor for the electromagnetic Lagrangian (\ref{densidadlag22}), we can use Eq.\,(\ref{amombelinfantee}) to define the Minkowski angular momentum tensor density as
\begin{equation}
m_{kl}{}^j:=x_k\Theta_l{}^j-x_l\Theta_k{}^j.\label{minkowskiangularmom}
\end{equation}
With the above definition, the spatial components of $m_{kl}{}^j$ are directly identified with the vectorial Minkowski angular momentum density $\bm{l}^{\rm M}:=\bm{x}\times\bm{\pi}^{\rm M}$ and the angular momentum flux ${\stackrel {\rm M}K}_a{}^b$, by
\begin{eqnarray}
m_{ab}{}^{0}&={}&-\epsilon_{abc}{\delta}^{cd}l_d^{\rm M}\label{amomiden1},\\
m_{ab}{}^c&={}&-\epsilon_{abd}{\delta}^{ce}{\stackrel {\rm M}K}_{e}{}^{d}.\label{amomiden2}
\end{eqnarray}
To obtain the balance equation for $m_{kl}{}^j$, we can either evaluate Eq.\,(\ref{beliamom3}) for the Lagrangian (\ref{densidadlag22}) or just take the 4-divergence of both sides of Eq.\,(\ref{minkowskiangularmom}) and use Eq.\,(\ref{divthetaonshell}). The result reads
\begin{equation}
\partial_{j} m_{kl}{}^{j}+{\stackrel {\rm J}{\cal T}}_{kl}+{\stackrel {\rm m}{\cal T}}_{kl}=0.\label{balance4momangular}
\end{equation} 
Here ${\stackrel {\rm J}{\cal T}}_{kl}$ is the antisymmetric Lorentz 4-torque density tensor, directly constructed from the Lorentz 4-force ${\cal F}^{\rm J}_i$:
\begin{equation}
{\stackrel {\rm J}{\cal T}}_{kl}:={}x_{k}{\cal F}_{l}^{\rm J}-x_{l}{\cal F}_{k}^{\rm J}.\label{tauJ}
\end{equation}
Its spatial components are expressed in terms of the Lorentz torque density $\bm{\tau}^{\rm J}:=\bm{x}\times\bm{f}^{\rm J}$ on the external charges and currents by
\begin{equation}
{\stackrel {\rm J}{\cal T}}_{ab}=-\epsilon_{abc}{\delta}^{cd}\tau_d^{\rm J}.\label{identificationtorque}
\end{equation}

Similarly to Eq.\,(\ref{effforce1}), we interpret ${\stackrel {\rm m}{\cal T}}_{kl}$  
as the \textit{effective material 4-torque density}, exerted by the electromagnetic field on the bound charges and currents of the material medium. Notice, however, that ${\stackrel {\rm m}{\cal T}}_{kl}$ is not constructed simply as the 4-torque of ${\cal F}^{\rm m}_i$, cf. Eq.\,(\ref{tauJ}), but it also depends on the antisymmetric part of the Minkowski tensor,
\begin{equation}
{\stackrel {\rm m}{\cal T}}_{kl}:={}x_{k}{\cal F}_{l}^{\rm m}-x_{l}{\cal F}_{k}^{\rm m}+2\Theta_{\left[kl\right]}\label{4taueffdef}.
\end{equation}
Using Eqs.\,(\ref{4const}), (\ref{minkowskitensor}) and (\ref{effforce1}) in Eq.\,(\ref{4taueffdef}), we obtain an explicit expression for the material 4-torque tensor ${\stackrel {\rm m}{\cal T}}_{kl}$ in terms of a general constitutive tensor:
\begin{eqnarray}
{\stackrel {\rm m}{\cal T}}_{kl}&={}&-\frac{1}{8}F^{ij}F^{mn}\left[x_{k}(\partial_{l}\chi_{ijmn})-x_{l}(\partial_{k}\chi_{ijmn})\right.\nonumber\\
&{}&+\,g_{ik}\chi_{ljmn}-g_{il}\chi_{kjmn}+g_{jk}\chi_{ilmn}-\,g_{jl}\chi_{ikmn}\nonumber\\
&{}&\left.+\,g_{mk}\chi_{ijln}-g_{ml}\chi_{ijkn}+\,g_{nk}\chi_{ijml}\right.\nonumber\\
&{}&-\left.g_{nl}\chi_{ijmk}\right].\label{taumexplicit}
\end{eqnarray}
If we compare Eq.\,(\ref{balance4momangular}) for $J^i=0$ with Eq.\,(\ref{beliamom3}), one can check that the above expression for ${\stackrel {\rm m}{\cal T}}_{kl}$ is actually defined in terms of the total angular momentum operator acting on the fourth rank constitutive tensor of the background medium:
\begin{equation}
{\stackrel {\rm m}{\cal T}}_{kl}=\frac{1}{8}\frac{\partial{\cal L}^{\rm e}}{\partial\chi^{ijmn}}(j_{kl})^{ijmn}{}_{opqr}\chi^{opqr}.\label{tauefflagg}
\end{equation}
Therefore, the interpretation of ${\stackrel {\rm m}{\cal T}}_{kl}$ is completely analogous to ${\cal F}^{\rm m}_i$, but this time related to the symmetry of the medium under spatial rotations and boosts (Lorentz transformations).  When the medium is isotropic, the corresponding $\chi^{ijkl}$ is invariant under 3-dimensional rotations, i.e. $(j_{ab})^{ijmn}{}_{opqr}\chi^{opqr}=0$, and thus the spatial components ${\stackrel {\rm m}{\cal T}}_{ab}$ in Eq.\,(\ref{balance4momangular}) vanish. As for ${\cal F}^{\rm m}_i$, this does not mean that in isotropic media light and matter do not microscopically interact, but they do it in such a way that there is no net macroscopic angular momentum transfer. As a consequence, the Minkowski angular momentum of light $\bm{l}^{\rm M}$, related to $m_{ab}{}^{0}$ by Eq.\,(\ref{amomiden1}), is a conserved quantity, provided $J^i=0$. On the contrary, if the medium presents local anisotropies, there will be a net material macroscopic torque ${\stackrel {\rm m}{\cal T}}_{ab}\neq 0$ and the angular momentum of light will change, for example by modifying its polarization. Therefore, the conservation of the Minkowski angular momentum of light in matter is not violated by the lack of symmetry with respect to all components of the Minkowski 4-tensor (\ref{minkowskitensor}). In fact, only the symmetry of the $3\times 3$ block of spatial components $\Theta_{ab}$ is important in Eq.\,(\ref{4taueffdef}) to make ${\stackrel {\rm m}{\cal T}}_{ab}=0$. 

On the other hand, the vanishing of the time-space components of the material 4-torque ${\stackrel {\rm m}{\cal T}}_{0a}$ is independently determined by the invariance of $\chi^{ijkl}$ under boosts. Nevertheless, as commented in Sec.\,\ref{opensystemmm}, the latter condition is fulfilled \emph{only} in the case that the medium is the vacuum, i.e. when $\chi^{ijkl}$ is constructed in terms of the Minkowski metric as in Eq.\,(\ref{constvacuum2}), since in that case the electromagnetic field alone form a closed system. In order to illustrate this, let us consider the second most simple medium, i.e. linear, homogeneous and isotropic medium at rest. In this case the constitutive tensor reads
\begin{equation}\label{constitutivehomoiso}
\chi^{ijkl}_{\rm iso,hom}:=\left(\gamma^{ik}\gamma^{jl}-\gamma^{il}\gamma^{jk}\right)/\mu_0\mu,
\end{equation}
constructed with the optical metric $\gamma^{ij}:={\rm diag}\left(n^2/c^2,-1,-1,-1\right)$ \cite{dielectricslab}. Here $n:=\sqrt{\varepsilon\mu}$ is the refractive index of the medium, with $\mu$ its relative permeability and $\varepsilon$ its relative permittivity. Inserting $\chi^{ijkl}_{\rm iso,hom}$ into Eq.\,(\ref{taumexplicit}), we can immediately verify that ${\stackrel {\rm m}{\cal T}}_{ab}=2\Theta_{[ab]}=0$, consistent with the isotropy symmetry, but the time-space components satisfy ${\stackrel {\rm m}{\cal T}}_{0a}=2\Theta_{[0a]}\propto n^2-1\neq 0$, for $n>1$. More generally, the very presence of a fixed background medium breaks the invariance of the open system under boosts, since an external agent is always needed to keep the medium fixed in its state of motion, in spite of the interaction of the medium with the electromagnetic field. Therefore, for any nontrivial medium, $\chi^{ijkl}$ is not invariant under boosts and specifically under the action of the spin operator:
\begin{equation}
2\Theta_{[0a]}=-\frac{1}{8}\frac{\partial{\cal L}^{\rm e}}{\partial\chi^{ijmn}}(s_{0a})^{ijmn}{}_{opqr}\chi^{opqr}\neq 0,\label{spacetimenonsymmetry}
\end{equation}
which implies that the Minkowski tensor is completely symmetric only in the trivial case of vacuum (\ref{symmetricminkowski}). From Eq.\,(\ref{spacetimenonsymmetry}) and the general identifications (\ref{identificationem}), we conclude that $\bm{\pi}^{\rm M}\neq\bm{S}^{\rm M}/c^2$, except for the case of the vacuum, where the system is actually a closed one. This is not surprising since the total BR tensor in a closed system is completely symmetric (\ref{BRsymmetricclosed}), however this is not necessarily the case for the total canonical one (\ref{nonsymmetrysigma}).

\subsubsection{Explicit Minkowski angular momentum balance equation}\label{minkowskiorbitalangularr}

For completeness, we explicitly write down the Minkowski angular momentum balance equation. If we evaluate Eq.\,(\ref{balance4momangular}) for $k,l=a,b$\ and use Eqs.\,(\ref{4const}), (\ref{effforce1}) and (\ref{4taueffdef}), together with the identifications (\ref{amomiden1}), (\ref{amomiden2}) and (\ref{identificationtorque}), we obtain
\begin{equation}
\frac{\partial l_a^{\rm M}}{\partial t}+\partial_b{\stackrel{\rm M}K}_a{}^b+\tau_a^{\rm m}+\tau_a^{\rm J}=0.\label{amom3}
\end{equation}
Here, the Minkowski angular momentum density $\bm{l}^{\rm M}$, the Minkowski angular momentum flux ${\stackrel {\rm M}K}_a{}^b$, the Lorentz torque on the external charges and currents $\bm{\tau}^{\rm J}$ and the effective material torque $\bm{\tau}^{\rm m}$ are explicitly given by:
\begin{eqnarray}
\bm{l}^{\rm M}&={}&\bm{x}\times\bm{\pi}^{\rm M}=(\bm{x}\cdot\bm{B})\bm{D}-(\bm{x}\cdot\bm{D})\bm{B},\label{emangularmom}\\
{\stackrel {\rm M}K}_a{}^b&={}&\epsilon_{acd}\delta^{de}x^cE_eD^b+\epsilon_{acd}\delta^{de}x^cH_eB^b\nonumber\\
&&-\frac{1}{2}\epsilon_{acd}\delta^{db}x^c(E_eD^e+H_eB^e),\\
\bm{\tau}^{\rm J}&={}&\bm{x}\times\bm{f}^{\rm J},\\
\tau_a^{\rm m}&=&-\frac{1}{2}\epsilon_{abc}x^b\delta^{cd}\left[D^e(\partial_d E_e)-E_e(\partial_d D^e)+B^e(\partial_d H_e)\right.\nonumber\\
&&\left.-H_e(\partial_d B^e)\right]+\epsilon_{abc}\delta^{bd}(E_dD^c+H_dB^c).
\end{eqnarray}
For the special case of a spatial volume $V$ without external charges and currents (hence $\tau_a^{\rm J}=0$) and an isotropic medium (hence $\tau_a^{\rm m}=0$), the integrated Minkowski angular momentum is a time independent quantity, provided the net angular momentum flux through the boundary $\partial V$ vanishes:
\begin{equation}
\bm{L}^{\rm M}=\int_V d^3x [(\bm{x}\cdot\bm{B})\bm{D}-(\bm{x}\cdot\bm{D})\bm{B}]
={\rm constant}.\label{minkowskiangularconserved}
\end{equation}

\subsection{Abraham tensor of light in matter}\label{abrahamtensorr}

The general structure of the Abraham tensor of the electromagnetic field in matter $\Omega_i{}^j$ is given, by definition, as the ``abrahamization'' of the Minkowski tensor $\Theta_i{}^j$ as \cite{obukhov1}
\begin{equation}\label{abraham}
\Omega_i{}^j:=\Theta_{i}{}^j-P_i{}^k\Theta_{[k}{}^{j]}-u_ku^j\Theta_{[i}{}^{k]}/c^2.
\end{equation}
Here $u^i:=(\gamma,\gamma \bm{v})$ is the 4-velocity field of the medium, with $\gamma:=\left(1-\bm{v}^2/c^2\right)^{-1/2}$ the usual Lorentz factor and 
\begin{equation}
P_i{}^k:=\delta_i^k-u_iu^k/c^2,\label{pprojector}
\end{equation}
the projector on the local rest frame of the medium. Inserting the general definition of the Minkowski tensor (\ref{minkowskitensor}) into Eq.\,(\ref{abraham}), we explicitly find
\begin{eqnarray}
\Omega_i{}^j&={}&\,{\frac 12}(F_{ik}H^{kj}+H_{ik}F^{kj})
+{\frac 14}\delta_i^jH^{kl}F_{kl}\nonumber\\
&{}&+\,{\frac 1{2c^2}}u_iu^l\left(F^{kj}H_{kl}-H^{kj}F_{kl}\right)\nonumber\\
&&+ {\frac 1{2c^2}}u^ju_l\left(F_{ki}H^{kl} - H_{ki}F^{kl}\right).\label{emtA}
\end{eqnarray}
It is important to stress here that Eq.\,({\ref{emtA}}) {\it is the definition of the Abraham tensor in any reference frame, for any medium with any constitutive relation (\ref{generalrelationss})}, which can even be non-local, non-linear, and dissipative, in general. The explicit form of the excitation tensor $H^{ij}$ will differ for different media, but the structure (\ref{emtA}) is maintained in the same way as the structure (\ref{minkowskitensor}). By construction, the Abraham tensor $\Omega_i{}^j$ is \emph{symmetric} at the cost of having the field $u^i$ explicitly in its definition. In contrast, the Minkowski tensor (\ref{minkowskitensor}) of a moving medium may depend on $u^i$ only implicitly through $H^{ij}$ and the constitutive relation (\ref{generalrelationss}). A second very important property of $\Omega_i{}^j$ is that for the special case of a medium at rest, i.e. when ${\stackrel {\circ}u}{}^i=(1,\bm{0})$, the Abraham energy density ${\stackrel {\circ}{\cal U}}{}^{\rm A}:={\stackrel {\circ}\Omega}_0{}^0$ and the Abraham energy flux density ${\stackrel {\circ}S}{}^a_{\rm A}:={\stackrel {\circ}\Omega}_0{}^a$ coincide with the corresponding Minkowski quantities in the same frame:
\begin{eqnarray}
{\stackrel {\circ}{\cal U}}{}^{\rm A}&={}&{\frac 12}(\bm{\stackrel {\circ}E}\cdot\bm{\stackrel {\circ}D}+\bm{\stackrel {\circ}H}\cdot\bm{\stackrel {\circ}B})={\stackrel {\circ}{\cal U}}{}^{\rm M},\\
\bm{\stackrel {\circ}S}{}^{\rm A}&={}&\bm{\stackrel {\circ}E}\times\bm{\stackrel {\circ}H}=\bm{\stackrel {\circ}S}{}^{\rm M}.
\end{eqnarray}
The latter property, together with the symmetry of $\Omega_i{}^j$ gives the famous expression for the Abraham momentum density ${\stackrel {\circ}\pi}{}_a^{\rm A}:=-{\stackrel {\circ}\Omega}_a{}^0$ for the medium at rest, which reads
\begin{equation}
\bm{\stackrel {\circ}\pi}{}^{\rm A}={}{\frac {\bm{\stackrel \circ S}{}^{\rm A}}{c^2}} = {\frac {\bm{\stackrel {\circ}E}\times\bm{\stackrel {\circ}H}}{c^2}}.\label{abrahammomdensity}
\end{equation}
For completeness, the remaining Abraham momentum flux ${\stackrel {\circ}p}{}^{\rm A}_a{}^b:=-{\stackrel {\circ}\Omega}_a{}^b$ is given by
\begin{eqnarray}
{\stackrel {\circ}p}{}^{\rm A}_a{}^b={\stackrel {\circ}p}{}^{\rm M}_{(a}{}^{b)}&={}&-\frac{1}{2}\left({\stackrel {\circ}E}_a{\stackrel {\circ}D}{}^b+{\stackrel {\circ}E}{}^b{\stackrel {\circ}D}_a+{\stackrel {\circ}H}_a{\stackrel {\circ}B}{}^b+{\stackrel {\circ}H}{}^b{\stackrel {\circ}B}_a\right)\nonumber\\
&{}&+\frac{1}{2}\delta{}^b_a\left({\stackrel {\circ}E}_c{\stackrel {\circ}D}{}^c+{\stackrel {\circ}H}_c{\stackrel {\circ}B}{}^c\right).
\end{eqnarray}
A word of caution is in order here. It is quite common to find in the literature the tacit treatment of Eq.\,(\ref{abrahammomdensity}) as a definition of the Abraham momentum, see for instance \cite{barnett2010,mansuripur,barnettloudon2,milonniboyd,baxterloudon,crenshaw}. In our opinion, this is an unfortunate practice. Unlike the Minkowski momentum density (\ref{momentM}) which always has the same form $\bm{\pi}^{\rm M}=\bm{D}\times\bm{B}$ for any medium in any reference frame, the Abraham momentum density has the form $\bm{{\stackrel {\circ}\pi}}{}^{\rm A} = (\bm{\stackrel {\circ}E}\times\bm{\stackrel {\circ}H})/c^2$ in the {\it rest frame} of matter only. This may easily lead to incorrect results, because in the framework of Special Relativity the expression (\ref{abrahammomdensity}) is no longer valid when the medium is moving. For example, Eq.\,(25) of Ref.~\cite{dielectricslab} shows the correct Abraham total momentum for a light pulse inside an homogeneous and isotropic medium, moving with constant velocity $\bm{v}$. Of course, there are certain situations when the velocity of the medium is so small, that the actual Abraham momentum can be very well approximated by the rest frame expression (\ref{abrahammomdensity}). However, one has to be very careful in order to avoid inconsistencies.

By taking the 4-divergence in both sides of Eq.\,(\ref{abraham}), we find a simple relation between the 4-divergences of the Minkowski and Abraham tensors:
\begin{equation}
\partial_j\Theta_i{}^j=\partial_{j}\Omega_i{}^{j}+{\cal F}_i^A,\label{relation}
\end{equation}
where ${\cal F}_i^A$ is the so-called Abraham 4-force density, defined in general by the following total 4-derivative:
\begin{eqnarray}
{\cal F}_{i}^A:&={}&\frac{1}{2}\partial_{j}\left[F_{ik}H^{kj}+\left(F_{ik}H^{kl}-H_{ik}F^{kl}\right)\frac{u^{j}u_{l}}{c^2}\right.\nonumber\\
&{}&\left.-H_{ik}F^{kj}-\left(F_{lk}H^{kj}-H_{lk}F^{kj}\right)\frac{u_{i}u^{l}}{{c^2}}\right].\label{afuerza}
\end{eqnarray}
In the important case that the medium is at rest, the above expression reduces to ${\stackrel \circ{\cal F}}{}_{i}^A=(0,-\bm{{\stackrel \circ f}}{}^A)$, where $\bm{{\stackrel \circ f}}{}^A$ is the famous Abraham force density in the rest frame:
\begin{eqnarray}
\bm{{\stackrel \circ f}}{}^A&=&\frac{\partial}{\partial t}\left(\bm{{\stackrel {\circ}D}}\times\bm{{\stackrel {\circ}B}}-\frac{1}{c^2}\bm{{\stackrel {\circ}E}}\times\bm{{\stackrel {\circ}H}}\right)\nonumber\\
&&+\frac{1}{2}\bm{\nabla}\times\left(\bm{{\stackrel {\circ}D}}\times\bm{{\stackrel {\circ}E}}+\bm{{\stackrel {\circ}B}}\times\bm{{\stackrel {\circ}H}}\right).\label{explicitabrahamrest}
\end{eqnarray}

Using Eq.\,(\ref{relation}), one can write the Minkowski energy-momentum balance equation (\ref{divthetaonshell}) in terms of the Abraham tensor and the Abraham 4-force (\ref{afuerza}), as it is done in \cite{obukhov1}. Thus,
\begin{equation}\label{abrahambalance}
\partial_{j}\Omega_i{}^{j}+{\cal F}_i^{\rm J}+{\cal F}_i^{\rm m}+{\cal F}_i^A=0,
\end{equation}
which in principle can be interpreted as the Abraham energy-momentum balance equation for a non-dynamical background medium, but in fact it is the same Minkowski balance equation with a rearrangement of terms. Therefore, the predictions from both balance equations are physically equivalent, but the corresponding interpretations may differ. Due to the extra {\it ad hoc} Abraham 4-force ${\cal F}_i^A$ in the balance equation (\ref{abrahambalance}), the Abraham tensor $\Omega_i{}^j$ is \emph{not conserved} when the background medium has symmetries such as spatial homogeneity, time independence or spatial isotropy, in contrast to the Minkowski tensor (\ref{minkowskitensor}). Even in the simplest nontrivial case of a light pulse propagating in an homogeneous, time-independent and isotropic medium at rest (\ref{constitutivehomoiso}), the Abraham force density (\ref{explicitabrahamrest}) is always nonzero,
\begin{equation}
\bm{{\stackrel \circ f}}{}^A_{\rm iso, hom}=\frac{(n^2-1)}{c^2}\frac{\partial}{\partial t}\left(\bm{{\stackrel {\circ}E}}\times\bm{{\stackrel {\circ}H}}\right)\neq 0,\label{isohomabrahamexp}
\end{equation}
implying in Eq.\,(\ref{abrahambalance}) that the Abraham momentum (\ref{abrahammomdensity}) is not conserved in that case. This predicted change in the content of Abraham momentum assigned to light is very impractical, since it does not have \textit{any observable consequence} in the propagation of the pulse. We know from Maxwell equations that in a simple medium light propagates immutably with constant amplitude and velocity $\bm{v}=(c/n)\bm{\hat{k}}$ and therefore there is no reason to assume the existence of any extra macroscopic force (\ref{isohomabrahamexp}) between light and matter. 

On the other hand, when assigning the Minkowski tensor (\ref{minkowskitensor}) for light in background media, the corresponding identifications of forces (\ref{4lorentzforce}) and (\ref{effforce1}) in the balance equation (\ref{divthetaonshell}) turn out to be completely consistent with the basic observations of the macroscopic propagation of light in matter, as we demonstrated in the previous subsections. Up to relocalizations (\ref{relT})-(\ref{relS}), the Minkowski tensor is the only one which is conserved under symmetry conditions of the medium and that is what makes it the \emph{most convenient choice for describing the energy-momentum content of light in matter as an open system.} This is in agreement with the conclusions of Ref.\,\cite{barnett2010}, however our result applies to the case of complex media and in contrast to \cite{barnett2010} it does not fundamentally invalidate the use of the Abraham tensor in any case.

Some authors have proposed and/or reported the measurement of the Abraham force  $\bm{{\stackrel \circ f}}{}^A$ as it appears in Eqs.\,(\ref{abrahambalance}) and (\ref{isohomabrahamexp}), using different setups \cite{james,wlw,ww,she,brevik2010,brevik2012,rikken,rikken2}. It is certainly correct to try to check the identical balance equations (\ref{minkowskitensor}) or (\ref{abrahambalance}), implied by the macroscopic Maxwell equations. However, it is impossible to discriminate between the Abraham and Minkowski momentum as the ``correct'' one for light in matter, since by definition both satisfy the same balance equation just interpreted in different ways.

\section{Electromagnetic field and dynamical medium as a closed system}\label{dynamicalmediumm}

\subsection{Lagrangian model for a dynamical medium interacting with light}

In the last two sections we considered the medium as a background on which the electromagnetic field propagates. Accordingly, the constitutive tensor $\chi^{ijkl}$ was treated as an external non-dynamical field coupled to the electromagnetic field. Now we will consistently extend our theory to include the dynamics of the material medium. We assume no other interaction with external electric charges and currents ($J^i=0$), so that the system composed of the electromagnetic field and the dynamical medium is closed. We follow the original approach of Penfield and Haus \cite{penfieldhaus1,penfieldhaus1967}, but in a more modern formulation as it is presented in \cite{obukhov1,ORR02}. 

Here we generalize previous dynamical models for complex media \cite{ORR02,mikura,obukhovhehl2007,leonhardt}, so that our resulting framework can be applied to any non-dissipative fluid medium with linear and non-dispersive optical properties. The central technical point is to express the constitutive tensor in terms of the dynamical matter fields,
\begin{equation}
\chi^{ijkl}=\chi^{ijkl}(u^i,\nu,\Psi^A).\label{dynamicalconstitutive}
\end{equation}
Here $u^{i}$ is the \textit{4-velocity field} of the medium and $\nu$ its \emph{particle number density}. The dynamical fields $\Psi^A(\bm{x},t)$ represent \textit{material variables} (of any tensor rank) different from $u^i$, $\nu$ that can be used to describe complex media with microstructure. An illustrative example of $\Psi^A$ is the 4-director field of a nematic liquid crystal as in Ref.\,\cite{ORR02}. We assume that $\chi^{ijkl}$ does \emph{not} depend on derivatives of the material variables, which only appear in the matter Lagrangian as shown below. 

We model the material medium as a fluid governed by the equations of relativistic hydrodynamics \cite{penfieldhaus1,obukhov1}. Therefore, the material variables that describe the translational dynamics of the medium should satisfy the following constraints:
\begin{eqnarray}
\partial_{i}(\nu u^{i})&={}&0,\label{continuidad}\\
u^{i}\partial_{i}s&={}&0,\label{entropia}\\
u^{i}\partial_{i}X&={}&0,\label{identidad}\\
u^{i}u_{i}&={}&c^2.\label{normalizacion}
\end{eqnarray}
Here $s$ is the \emph{entropy density} and $X$ is the so-called \emph{identity (Lin) coordinate}. The first Eq.\,(\ref{continuidad}) expresses the particle number conservation. The second Eq.\,(\ref{entropia}) assumes entropy conservation along the streamlines of the fluid so that only reversible processes are allowed (\emph{adiabatic} or \emph{isentropic} fluid). The entropy along different streamlines does not have to be necessarily the same and thus one introduces the Lin coordinate $X$ to identify the particles from different streamlines \cite{penfieldhaus1,lin,ray72}. The identity variable $X$ has to be conserved on each streamline, hence Eq.\,(\ref{identidad}). The last Eq.\,(\ref{normalizacion}) is the usual relativistic normalization of the 4-velocity field.

The total Lagrangian of the coupled system of electromagnetic field and matter reads 
\begin{equation}
{\cal L}={\cal L}^{\rm e}+{\cal L}^{\rm m},\label{materialmodellag}
\end{equation}
where 
${\cal L}^{\rm e}$ is the electromagnetic Lagrangian (\ref{densidadlag22}) for $F_{ij}=\partial_iA_j-\partial_jA_i$, with a constitutive tensor of the form (\ref{dynamicalconstitutive}) and ${\cal L}^{\rm m}$ is the dynamical matter Lagrangian given explicitly by
\begin{eqnarray}
{\cal L}^{\rm m}&=&-\,\rho(\nu,s)+\Lambda_0(u^{i}u_{i}-c^2)-\nu u^{i}\partial_{i}\Lambda_1\nonumber\\
&&+\Lambda_2 u^{i}\partial_{i}s+\Lambda_3 u^{i}\partial_{i}X+{\cal L}^{\rm a}.\label{Lmat}
\end{eqnarray}
Here $\rho(\nu,s)$ is the thermodynamic \emph{internal energy density} of the fluid, $\Lambda_I$, with $I=0,...,3$ are Lagrange multipliers that we use to impose the constraints (\ref{continuidad})-(\ref{normalizacion}) and 
\begin{equation}
{\cal L}^{\rm a}={\cal L}^{\rm a}(\nu,\partial_i\nu,u^i,\partial_ju^i,\Psi^A,\partial_i\Psi^A)
\end{equation}
is the material Lagrangian that includes kinetic and potential terms for the matter fields $\nu$, $u^i$, $\Psi^A$ and their derivatives. When ${\cal L}^{\rm a}=0$, the Lagrangian (\ref{Lmat}) describes a dynamical isotropic ideal fluid \cite{obukhov1}, the elements of which do not have internal structure. Therefore, the anisotropic contribution  ${\cal L}^{\rm a}$ in the fluid Lagrangian describes the possible complex nature of material elements of the medium, thus generalizing the dynamical models of matter with microstructure \cite{WR,halb,Kopc,OK,OT}. For instance, in Ref.\,\cite{ORR02} we use a particular choice  of ${\cal L}^{\rm a}$ to describe the dynamics of a relativistic liquid crystal medium.

\subsection{Equations of motion}

Following the lines of Refs.\,\cite{obukhov1,ORR02}, we now will derive the equations of motion (\ref{EL}) of the total Lagrangian (\ref{materialmodellag}) for the dynamical fields of the closed system. We will subsequently use them to obtain the explicit expression for the BR tensor of the matter subsystem in Sec.\,\ref{matteremtensor}. 

The Euler-Lagrange equations for the electromagnetic potential $A_i$ are the usual Maxwell equations (\ref{Maxwellsequations}) and the ones corresponding to the Lagrange multipliers $\Lambda_I$ give, by construction, the dynamical constraints (\ref{continuidad})-(\ref{normalizacion}). In addition, the variation of $\cal{L}$ with respect to the material variables $\Psi^A$, $\nu$ and $u^i$ yields, respectively,  
\begin{eqnarray}
&&\frac{\delta{\cal L}^{\rm a}}{\delta\Psi^A} + \frac{\partial{\cal L}^{\rm e}}{\partial\Psi^A} = 0,\label{eqPsi}\\
&& -\nu u^i\partial_i\Lambda_1 = \nu\frac{\partial\rho}{\partial\nu} - \nu \frac{\delta{\cal L}^{\rm a}}{\delta\nu}-\nu\frac{\partial{\cal L}^{\rm e}}{\partial\nu},\label{eqnu}\\
&&\nu\partial_i\Lambda_1\!-\!\Lambda_2\partial_i s\!-\!\Lambda_3\partial_i X \!=\! 2\Lambda_0u_i\!+\!\frac{\partial{\cal L}^{\rm e}}{\partial u^i}\!+\!\frac{\delta{\cal L}^{\rm a}}{\delta u^i}.\label{equ}
\end{eqnarray}
Notice that Eq.\,(\ref{eqPsi}) includes all the constraints and dynamical equations that the general material variables $\Psi^A$ must satisfy in the specific model for the medium.

We introduce the temperature $T$ and the pressure $p$ of the material medium via the Gibbs law of thermodynamics, which in this case can be written as \cite{obukhov1,ORR02}
\begin{equation}
d\left(\frac{\rho}{\nu}\right)=Tds-p d\left(\frac{1}{\nu}\right).\label{gibbs}
\end{equation}
From Eq.\,(\ref{gibbs}) it is straightforward to show that $\nu(\partial\rho/\partial\nu)=\rho+p$. Using the latter relation together with the equations of motion (\ref{continuidad})-(\ref{normalizacion}) and (\ref{eqnu}) in Eq.\,(\ref{Lmat}), we see that ${\cal L}^{\rm m}$ can be written very simply as
\begin{equation}
{\cal L}^{\rm m}=\tilde{p}+{\cal L}^{\rm a},\label{simplelmat}
\end{equation}
where we defined the \emph{effective pressure} $\tilde{p}$ as
\begin{equation}\label{peff}
\tilde{p}:=p-\nu{\frac{\partial{\cal L}^{\rm e}}{\partial\nu}}-\nu{\frac{\delta{\cal L}^{\rm a}}{\delta\nu}}.
\end{equation}

\subsection{Canonical and Belinfante-Rosenfeld tensors for the matter subsystem}\label{matteremtensor}

In Secs.\,\ref{canonicalemsection}-\ref{subminkbeli} we already derived the canonical and BR tensors for the electromagnetic field subsystem. To derive now the canonical energy-momentum tensor for matter ${\stackrel{\rm m} \Sigma}_i{}^j$, we insert the matter Lagrangian (\ref{Lmat}) into the general definition (\ref{canonicalemtensor}). Thereby, we obtain
\begin{equation}
{\stackrel{\rm m}\Sigma}{}_i{}^j=-2\Lambda_0u_iu^j-u^j\left(\frac{\delta{\cal L}^{\rm e}}{\delta u^i}+\frac{\delta{\cal L}^{\rm a}}{\delta u^i}\right)-\delta_i^j\,\tilde{p}+{\stackrel{\rm a}\Sigma}_i{}^j,\label{cemm1}
\end{equation}
where we also used Eqs.\,(\ref{eqnu}) and (\ref{peff}), and the anisotropic energy-momentum tensor $\stackrel{\rm a}{\Sigma}_i{}^j$ is defined as
\begin{equation}
{\stackrel{\rm a}\Sigma}_i{}^j:=\frac{\partial{\cal L}^{\rm a}}{\partial
(\partial_j\Psi^A)}\,\partial_i\Psi^A+\frac{\partial{\cal L}^{\rm a}}{\partial
(\partial_ju^k)}\,\partial_iu^k+\frac{\partial{\cal L}^{\rm a}}{\partial
(\partial_j\nu)}\,\partial_i\nu-\delta_i^j\,{\cal L}^{\rm a}.
\end{equation}
Contracting Eq.\,(\ref{equ}) with $u^i$ and then using the other equations of motion (\ref{entropia})-(\ref{normalizacion}) and (\ref{eqnu}), one obtains $-2\Lambda_0=(\rho+\tilde{p})/c^2+(u^i/c^2)\left(\partial{\cal L}^{\rm e}/\partial u^i+\delta{\cal L}^{\rm a}/\delta u^i\right)$. Inserting this into Eq.\,(\ref{cemm1}), we eliminate all the Lagrange multipliers and recast the matter canonical tensor into
\begin{equation}
{\stackrel{\rm m}\Sigma}{}_i{}^j=\rho\frac{u_iu^j}{c^2}-P_i{}^j\,\tilde{p}+{\stackrel{\rm a}\Sigma}_i{}^j-u^jP_i{}^k\left(\frac{\partial{\cal L}^{\rm e}}{\partial u^k}+\frac{\delta{\cal L}^{\rm a}}{\delta u^k} \right),\label{cemm2}
\end{equation}
where the projector $P_i{}^j$ was defined in Eq.\,(\ref{pprojector}).

To construct the BR tensor for matter ${\stackrel {\rm m}\sigma}_i{}^j$, we need to know the corresponding spin density and its derivatives. Inserting Eq.\,(\ref{Lmat}) into Eq.\,(\ref{spindef}), we see that the matter spin ${\stackrel {\rm m}S}_{ij}{}^k$ has only contributions from ${\cal L}^{\rm a}$:
\begin{equation}
{\stackrel {\rm m}S}_{ij}{}^k=\frac{\partial{\cal L}^{\rm a}}{\partial(\partial_k\Psi^A)}(s_{ij})^A{}_B\Psi^B+2\frac{\partial{\cal L}^{\rm a}}{\partial(\partial_k u^{[i})}u_{j]},
\end{equation}
where $(s_{ij})^A{}_B$ are the Lorentz generators for the general matter fields $\Psi^A$.
Using the angular momentum identity (\ref{angularidentity}), the Euler-Lagrange equation for the closed system, $\delta{\cal L}/\delta u^i=\delta{\cal L}^{\rm m}/\delta u^i+\partial{\cal L}^{\rm e}/\partial u^i=0$, and the Lorentz generator of a 4-vector, $(s_{ij})^k{}_l:=\delta^k_ig_{jl}-\delta^k_jg_{il}$, we can obtain a useful expression for the contraction $\partial_k{\stackrel {\rm m}S}_{ij}{}^k$:
\begin{equation}
\partial_k{\stackrel {\rm m}S}_{ij}{}^k=2{\stackrel {\rm m}\Sigma}_{[ij]}+2\frac{\partial{\cal L}^{\rm e}}{\partial u^{[i}}u_{j]}-\frac{\delta{\cal L}^{\rm a}}{\delta \Psi^A}(s_{ij})^A{}_B\Psi^B.\label{Sderivate}
\end{equation}
Finally, if we recall the general definition of the BR tensor (\ref{BR}) and use Eqs.\,(\ref{cemm2})-(\ref{Sderivate}), we find the BR tensor for the matter subsystem,
\begin{eqnarray}
{\stackrel{\rm m}\sigma}_{i}{}^{j}&={}&\rho\frac{u_iu^j}{c^2}-P_i{}^j\,\tilde{p}-u^{j}P_{i}{}^k\frac{\partial{\cal L}^{\rm e}}{\partial u^k}-u^{(j}P_{i)}{}^k\frac{\delta{\cal L}^{\rm a}}{\delta u^k}+{\stackrel{\rm a}\Sigma}_{(i}{}^{j)}\nonumber\\
&+&\frac{1}{2}\frac{\delta{\cal L}^{\rm a}}{\delta \Psi^A}(s_{i}{}^j)^A{}_B\Psi^B+\frac{1}{2}\partial_{k}({\stackrel {\rm m}S}{}^{jk}{}_{i}+{\stackrel {\rm m}S}_{i}{}^{kj}).\label{matterbelinfante}
\end{eqnarray}
The BR tensor for matter ${\stackrel{\rm m}\sigma}_{i}{}^{j}$ depends not only on the material fields, but also on the electromagnetic field $F_{ij}$ via the term $-u^{j}P_{i}{}^k(\partial{\cal L}^{\rm e}/\partial u^k)$ and implicitly via the effective pressure (\ref{peff}) due to the electro- and magnetostriction term $\nu(\partial{\cal L}^{\rm e}/\partial \nu)$.  

\subsection{Total Belinfante-Rosenfeld tensor of the closed system}

We found in Sec.\,\ref{subminkbeli} that the Minkowski tensor (\ref{minkowskitensor}) is actually the BR tensor of the electromagnetic field subsystem ${\stackrel {\rm e}\sigma}_i{}^j=\Theta_i{}^j$, when $J^i=0$. The total BR tensor ${\stackrel {\rm t}\sigma}_i{}^j$ of the closed system (\ref{materialmodellag}) is obtained as the sum of the BR tensor of the material medium (\ref{matterbelinfante}) and the Minkowski tensor for the electromagnetic field (\ref{minkowskitensor}):
\begin{equation}
{\stackrel {\rm t}\sigma}_i{}^j:=\Theta_i{}^j+{\stackrel {\rm m}\sigma}_i{}^j.\label{totalBRmink}
\end{equation}
The total BR tensor (\ref{totalBRmink}) is the physically relevant quantity to describe the energy, momentum and angular momentum content of the closed system, since it always satisfies a conservation equation, together with the associated angular momentum ${\stackrel {\rm t}l}_{kl}{}^j:=x_k{\stackrel {\rm t}\sigma}_l{}^j-x_l{\stackrel {\rm t}\sigma}_k{}^j$ (\ref{conservationclose}):
\begin{equation}
\partial_j{\stackrel {\rm t}\sigma}_i{}^j=0,\qquad \partial_j{\stackrel {\rm t}l}_{kl}{}^j=0.\label{BRconservationeqs}
\end{equation}
In addition, the total BR tensor is, by construction, fully symmetric ${\stackrel {\rm t}\sigma}_{[ij]}=0$, c.f. Eq.\,(\ref{BRsymmetricclosed}), and depends on all the dynamical fields of the closed system.

\section{General Abraham decomposition of the total Belinfante-Rosenfeld tensor}\label{TBR}

In this section we explicitly find a general decomposition of the total BR tensor ${\stackrel {\rm t}\sigma}_i{}^j$ in terms of the Abraham tensor $\Omega_i{}^j$, which explains why the latter turns out to be very convenient to describe the energy-momentum of light in dynamical matter. To construct the Abraham tensor $\Omega_i{}^j$ as the ``abrahamization'' (\ref{abraham}) of the Minkowski tensor $\Theta_i{}^j$, we first need to compute its antisymmetric part $\Theta_{[ij]}$. This can be obtained from the angular momentum identity (\ref{beliarelation3}) applied only to the electromagnetic Lagrangian (\ref{densidadlag22}). Using also the equations of motion (\ref{eqPsi}) and the Lorentz generator $(s_{ij})^k{}_l=\delta^k_ig_{jl}-\delta^k_jg_{il}$, we obtain
\begin{equation}
\Theta_{[ij]}={}\frac{\partial{\cal L}^{\rm e}}{\partial u^{[i}}u_{j]}-\frac{1}{2}\frac{\delta{\cal L}^{\rm a}}{\delta\Psi^A}(s_{ij})^A{}_B\Psi^B.\label{antimink}
\end{equation}
Then, inserting Eq.\,(\ref{antimink}) into Eq.\,(\ref{abraham}), we find a relation between the Minkowski and Abraham tensors in material media with the general linear constitutive law (\ref{dynamicalconstitutive}):
\begin{eqnarray}
\Omega_i{}^j&=&\Theta_i{}^j - u^jP_i{}^k\frac{\partial{\cal L}^{\rm e}}{\partial u^{k}} +\frac{1}{2}\frac{\delta{\cal L}^{\rm a}}{\delta  \Psi^A}(s_i{}^j)^A{}_B\Psi^B\nonumber\\
&&+{\frac 1{c^2}}u^ku^{(j}\frac{\delta{\cal L}^{\rm a}}{\delta \Psi^A}(s_{i)k})^A{}_B\Psi^B.\label{miab}
\end{eqnarray}
In particular, when all the matter fields $\Psi^A$ are \textit{scalars} ($u^i$ is not included there), the last two terms in Eq.\,(\ref{miab}) vanish, and we recover the simple relation between $\Theta_i{}^j$ and $\Omega_i{}^j$ derived for simple isotropic media in \cite{obukhov1}, cf. Eq.\,(106).

If we now use Eq.\,(\ref{miab}) in Eq.\,(\ref{totalBRmink}), we can re-express the Minkowski tensor in terms of the Abraham tensor and discover a crucial result: Except for the effective pressure (\ref{peff}), \emph{all the terms of ${\stackrel {\rm t}\sigma}_i{}^j$ depending explicitly on the electromagnetic field $F_{ij}$ are contained in the definition of $\Omega_{i}{}^j$.} Therefore, alternatively as in Eq.\,(\ref{totalBRmink}), the total BR tensor can be conveniently decomposed as:
\begin{equation}
{\stackrel {\rm t}\sigma}_i{}^j={}\Omega_i{}^j+{\stackrel{\rm m}\kappa}_i{}^j,\label{sigmatot}
\end{equation}
where ${\stackrel {\rm m}\kappa}_i{}^j:={\stackrel {\rm t}\sigma}_i{}^j-\Omega_i{}^j$ depends only on matter fields (except by $\tilde{p}$) and it is explicitly given by
\begin{eqnarray}
{\stackrel{\rm m}\kappa}_i{}^j&={}&\rho\frac{u_iu^j}{c^2}-P_i{}^j\,\tilde{p}+{\stackrel{\rm a}\Sigma}_{(i}{}^{j)}+\frac{1}{2}\partial_{k}({\stackrel {\rm m}S}{}^{jk}{}_{i}+{\stackrel {\rm m}S}_{i}{}^{kj})\nonumber\\
&{}& -u^{(j}P_{i)}{}^k\frac{\delta{\cal L}^{\rm a}}{\delta u^k}-\frac{u^ku^{(j}}{c^2}\frac{\delta{\cal L}^{\rm a}}{\delta \Psi^A}(s_{i)k})^A{}_B\Psi^B.\label{matterabrahamdef}
\end{eqnarray}
Generalizing \cite{barnett2010,hindsbarnett}, we call ${\stackrel {\rm m}\kappa}_i{}^j$ the \textit{kinetic energy-momentum tensor of matter} in order to distinguish it from the ``canonical'' BR tensor ${\stackrel {\rm m}\sigma}_i{}^j$ in Eq.\,(\ref{matterbelinfante}). In most practical cases, the magneto- and electrostriction effects can be neglected and thus the kinetic matter tensor ${\stackrel{\rm m}\kappa}_i{}^j$ depends only on the material fields of the system.

Equations (\ref{totalBRmink}) and (\ref{sigmatot}) are of course physically equivalent, but the ``Abraham decomposition'' (\ref{sigmatot}) has important interpretational advantages as compared to the ``Minkowski decomposition'' (\ref{totalBRmink}). First, in Eq.\,(\ref{sigmatot}) both tensors are separately symmetric, whereas for the Minkowski decomposition this is only true for total sum ${\stackrel {\rm t}\sigma}_i{}^j$. Second and more importantly, if we use the Abraham tensor $\Omega_i{}^j$ to describe the energy-momentum content of light in dynamical matter, the conservation equation for the total system can be conveniently written as
\begin{equation}
\partial_j\Omega_i{}^j=-\partial_j{\stackrel{\rm m}\kappa}_i{}^j.\label{abrahamconservation}
\end{equation}
Since ${\stackrel{\rm m}\kappa}_i{}^j$ contains only matter quantities and $\Omega_i{}^j$ contains all the electromagnetic terms, the conservation equation (\ref{abrahamconservation}) can be understood as energy and momentum transfers between two ``almost decoupled'' subsystems. The term $\partial_j\Omega_i{}^j$ describes the change of Abraham energy-momentum associated to the electromagnetic field, which is compensated by a the corresponding change of kinetic energy-momentum $\partial_j{\stackrel{\rm m}\kappa}_i{}^j$ associated to the dynamical medium. A similar interpretation in the Minkowski conservation equation $\partial_j\Theta_i{}^j=-\partial_j{\stackrel {\rm m}\sigma}_i{}^j$ is not possible, since the BR tensor of matter ${\stackrel {\rm m}\sigma}_i{}^j$ always contains the electromagnetic field $F_{ij}$.

The most simple example one could think of to verify the general decomposition (\ref{sigmatot}) is an isotropic medium, where ${\cal L}^{\rm a}=0$. In that case, as shown explicitly in Eq.\,(95) of Ref.\,\cite{obukhov1}, the kinetic tensor turns out to be the energy-momentum tensor of an ideal fluid with effective pressure: ${\stackrel{\rm m}\kappa}{}_i{}^j = \rho u_iu^j/c^2-P_i{}^j \tilde{p}$. A less trivial medium is a liquid crystal with anisotropic uniaxial electromagnetic properties, whose relativistic Lagrangian theory was developed in Ref.\,\cite{ORR02}. For this physically relevant case, we derived the corresponding total BR tensor and the Abraham tensor, which allowed us to explicitly verify the general Abraham decomposition (\ref{sigmatot}). The explicit results are quite bulky and can be found in \ref{exampleappendix}.

\section{Conclusions and discussion}\label{summary}

Developing a Lagrangian formalism in the framework of the classical field theory, we have revisited the problem of the correct definition and interpretation of energy-momentum tensors for the electromagnetic field in matter. Our results are very general in the sense that they apply to any linear, non-dispersive, non-dissipative and possibly moving media. We demonstrated that the general definition of Belinfante-Rosenfeld tensor provides a way to understand the meaning of previously considered energy-momentum tensors, among which the Minkowski and Abraham tensors are the most prominent ones. We have found that in order to avoid confusions in determining a consistent energy-momentum tensor for light in a medium, it is crucial to distinguish whether the system under consideration is open or closed. In each case we highlight the qualitatively different properties of the tensors describing the corresponding physical system.

When the material medium is assumed to be a non-dynamical background for the propagation of light (as in the second example of Ref.\,\cite{barnett2010}), then the electromagnetic field should be treated as an open system. In this case, the Minkowski tensor is naturally interpreted as the energy-momentum tensor of light in matter, since its conservation is determined by the symmetries of the background medium, such as spatial homogeneity, time independence or spatial isotropy. The macroscopic Maxwell equations imply balance equations for the Minkowski energy-momentum (\ref{divthetaonshell}) and Minkowski angular momentum (\ref{balance4momangular}), where we identified a macroscopic material 4-force (\ref{effforce1}) and 4-torque (\ref{taumexplicit}), that are consistent with basic observations of light propagation in media. We additionally demonstrated that the asymmetry of the Minkowski tensor is actually necessary for the consistent description of the interaction of light with a general linear medium. In particular, the Minkowski angular momentum is conserved in the case of isotropic media at rest, c.f.  Eq.\,(\ref{minkowskiangularconserved}), even though the Minkowski tensor is not fully symmetric. In view of its close relation to the canonical energy-momentum tensor (\ref{canonicalem}), it is natural to interpret the Minkowski momentum as the ``canonical'' momentum of light, as it is also discussed in Refs.\,\cite{obukhov1,barnett2010}. On the other hand, we found that the Abraham tensor is not particularly useful for the case of a background medium. It satisfies the same balance equations (\ref{abrahambalance}), but in contrast to the Minkowski tensor, it is not conserved when the medium has symmetries. This is usually described by an {\it ad hoc} introduction of the macroscopic Abraham force, which has no observable consequence in the propagation of light in homogeneous media. As a matter of fact, the attempts to measure this Abraham force \cite{james,wlw,ww,she,brevik2010,brevik2012,rikken,rikken2} can only probe the validity of the balance equations (\ref{divthetaonshell}), or equivalently (\ref{abrahambalance}), but it cannot discriminate between the Abraham and Minkowski tensors. 

A qualitatively different situation arises when the medium can move due to its interaction with the electromagnetic field, since in that case light and material medium form a closed system. To describe the dynamics of the medium in the Lagrangian framework, we model the matter as a relativistic non-dissipative fluid with a linear non-dispersive constitutive law (\ref{dynamicalconstitutive}) that can depend on material fields $\Psi^A$. By extending the original approach of Penfield and Haus \cite{penfieldhaus1,penfieldhaus1967} we derived an explicit expression for the BR tensor of matter (\ref{matterbelinfante}) which, when added to the Minkowski tensor, forms the conserved and symmetric total BR tensor of the closed system (\ref{totalBRmink})-(\ref{BRconservationeqs}). The total canonical tensor is equally valid and conserved, but is not symmetric (\ref{nonsymmetrysigma}). Since the electromagnetic and material variables are coupled through the field equations (\ref{Maxwellsequations}) and (\ref{eqPsi}), only the total energy-momentum tensors have clear physical meaning, a fact that has been pointed out by many researchers \cite{penfieldhaus1,saldanha,pfeifer,robinson,mikura,israel,brevik}. A decomposition of the total tensor into a light and matter parts is quite arbitrary since one cannot independently measure them in an experiment, however some specific decompositions can  more convenient or useful than others \cite{saldanha}.

The total BR tensor ${\stackrel {\rm t}\sigma}_i{}^j$ turns out to be particularly convenient in contrast to the total canonical one, since for general linear media it can be decomposed in two important ways. Namely, it can be represented: a) by definition, as the sum of the Minkowski tensor of light $\Theta_i{}^j$ and the ``canonical'' BR tensor of the medium ${\stackrel {\rm m}\sigma}_i{}^j$ (\ref{totalBRmink}) or, alternatively, b) as a sum of the Abraham tensor of light $\Omega_i{}^j$ and the ``kinetic'' tensor of the medium ${\stackrel {\rm m}\kappa}_i{}^j$ (\ref{sigmatot}). The Abraham decomposition ${\stackrel {\rm t}\sigma}_i{}^j=\Omega_i{}^j+{\stackrel{\rm m}\kappa}_i{}^j$ is particularly convenient to describe light in dynamical media since in that case the kinetic tensor ${\stackrel{\rm m}\kappa}_i{}^j$ assigned to matter contains only material dynamical fields and the Abraham tensor $\Omega_i{}^j$ assigned to light contains all the electromagnetic terms with $F_{ij}$ (except by the usually negligible electro- and magnetostriction effects). This seems to be a non-trivial result, since there is no \emph{a priori} reason to expect that the part of the total BR tensor that contains the electromagnetic field would exactly reproduce the structure derived from the general algebraic definition (\ref{abraham}) of the Abraham tensor \textit{for all types of non-dispersive and non-dissipative linear media}. In addition, ${\stackrel{\rm m}\kappa}_i{}^j$ and $\Omega_i{}^j$ are both separately symmetric, whereas for the Minkowski decomposition that is only true for the total sum ${\stackrel {\rm t}\sigma}_i{}^j=\Theta_i{}^j+{\stackrel{\rm m}\sigma}_i{}^j$. With all these nice properties of the Abraham decomposition, the interaction between the electromagnetic field and \emph{dynamical} material medium can be described in a very simple way as energy and momentum transfers between two ``almost decoupled'' subsystems (\ref{abrahamconservation}). As a result, the change of the Abraham energy, momentum, and angular momentum assigned to the field is exactly compensated by a change of the  energy, momentum, and angular momentum of the medium, which has the \textit{same form} as if the electromagnetic field was not present. In this sense, the Abraham momentum can be indeed considered as the ``kinetic'' momentum of the field in analogy to Refs.\,\cite{barnett2010,hindsbarnett}.

In Refs. \cite{barnett2010,barnettloudon2} Barnett and Loudon obtain an equation similar to Eqs.\,(\ref{totalBRmink}) and (\ref{sigmatot}), but for the special case of light propagating inside an homogeneous and isotropic medium \emph{at rest}. By considering some very illustrative examples, they correctly identify the Abraham momentum as the kinetic momentum of light and the Minkowski momentum as the canonical momentum in isotropic simple media. Based on classical field theory, here we extended their analysis to the case of complex anisotropic linear media in motion and we provide a better theoretical basis to understand their resolution \cite{barnett2010,barnettloudon2}. On the other hand, Barnett and Loudon did not consider the key point of clearly distinguishing the cases when the medium is assumed to be fixed or dynamical (i.e., when the system is open or closed) and therefore they incorrectly concluded that the Abraham and Minkowski momenta are two mutually exclusive alternatives for the description of light in media. An explicit computation and discussion related to this point can be found in \cite{dielectricslab}.

Within the Lagrangian framework developed in this work, we give the first principles arguments to understand the long-standing Abraham-Minkowski controversy for light in material media. The classical theory of electrodynamics in macroscopic media \cite{jackson,landau,LL} is (and has always been) perfectly consistent, although there are certain subtleties in the interpretations that one has to consider in order to avoid confusions. A next challenge would be to test explicitly this general framework in more specific and new situations, both theoretically and experimentally. An interesting question is whether it is possible to generalize our results to even more complex media (for instance, dissipative \cite{fukuma1,fukuma2}, dispersive \cite{Furutsu,philbin,philbin2}, non-local \cite{castaldi,mashhoon1,mashhoon2,mashhoon3} or non-linear media \cite{ogr,lorenci,pereiraklippert}) in order to find consistent energy-momentum tensors for each case.

%%%%%%%%%%%%%%%%%%%%%%%%%%%%%%%%%%%%%%%%%%%%%%%%%
\section*{Acknowledgments}
%%%%%%%%%%%%%%%%%%%%%%%%%%%%%%%%%%%%%%%%%%%%%%%%%

T.R. acknowledges financial support from BECAS CHILE scholarship program. 
For Y.N.O. this work was partly supported by the German-Israeli Foundation for
Scientific Research and Development (GIF), Research Grant No.\ 1078-107.14/2009. 

\appendix

\section{Uniaxial anisotropic medium interacting with light}\label{exampleappendix}

As a particular example that illustrates how this general BR-Abraham relation (\ref{sigmatot}) works, we will consider the special case of a nematic liquid crystal, for which a relativistic fluid model has been recently developed in Ref.\,\cite{ORR02}. In that paper, the total canonical energy-momentum tensor $\Sigma_i{}^j$ of the closed system composed of electromagnetic field and an uniaxial anisotropic medium was explicitly derived.  $\Sigma_i{}^j$ is not symmetric, cf. Eq.\,(9.11) of Ref.\,\cite{ORR02}, even though the total system is closed. The reason is that the total spin density of the system $S_{ij}{}^k$ does not vanish (\ref{nonsymmetrysigma}). 

Now we will complete the discussion of this case by explicitly computing the corresponding total BR energy-momentum tensor (\ref{BR}) and the Abraham tensor (\ref{emtA}), so that we can at the end verify the relation (\ref{sigmatot}) in a concrete system of physical importance. The liquid crystal medium can be described using the matter Lagrangian (\ref{materialmodellag}), where the anisotropic Lagrangian ${\cal L}^{\rm a}$ has the following particular form \cite{ORR02}:
\begin{eqnarray}
{\cal L}^{\rm a}&=&-\frac{1}{2}J\nu\omega^{i}\omega_{i}+\Lambda_4(N^{i}N_{i}+1)+\Lambda_5 u^{i}N_{i}\nonumber\\ &-&\frac{1}{2}K_1\left(\partial_{i}N^{i}\right)^2-\frac{1}{2}K_2\left(\epsilon^{ijk}N_{i}\partial_{j}N_{k}\right)^2\nonumber\\
&&+\frac{1}{2}K_3\left(\epsilon_{ijk}N^{j}\epsilon^{kln}\partial_{l}N_{n}\right)^2.\label{Lfluid1}
\end{eqnarray}
Here $\Psi^A = \{N^i, \Lambda_4, \Lambda_5\}$, where $N^i$ is the {\it director} 4-vector. It has a unit length, is spacelike and orthogonal to the velocity $u^i$; $\omega^{i}:=\epsilon^{ijk}N_{j}\dot{N}_{k}$ is the relativistic 4-dimensional angular velocity. The convective derivative reads $\dot{N}^{i}:=u^{j}\partial_{j}N^{i}$ and the 3-dimensional Levi-Civita symbol is defined as $\epsilon_{ijk}:=\eta_{ijkl}u^{l}/c$, with the 4-dimensional Levi-Civita tensor defined such that $\eta_{0123}:=c$. $K_1, K_2$ and $K_3$ are the Frank elastic constants, describing the internal deformation interaction of the liquid crystal and $J$ is the moment of inertia of matter elements of the fluid. 

On the other hand, the electromagnetic properties of the uniaxial dielectric and diamagnetic medium can be described by the permittivities $\varepsilon_{||}$ along the symmetry axis and $\varepsilon_{\perp}=\varepsilon$ perpendicular to the axis, and similarly by the permeabilities $\mu_{||}$ along the axis and $\mu_{\perp}=\mu$ perpendicular to the axis. The electromagnetic Lagrangian (\ref{densidadlag22}) is then specified by the corresponding constitutive tensor $\chi^{ijkl}$ of an uniaxial medium, which reads \cite{ORR02}
\begin{widetext}
\begin{align}
\chi^{ijkl}=&\,\frac{\left(\mu^{-1}_{\perp}+\Delta\mu^{-1}\right)}{\mu_0}\left(g^{ik}g^{jl}-g^{il}g^{jk}\right)-\frac{\left(\Delta\varepsilon+\Delta\mu^{-1}\right)}{\mu_0c^2}\left(u^{i}u^{k}N^{j}N^{l}-u^{i}u^{l}N^{j}N^{k}+u^{j}u^{l}N^{i}N^{k}-u^{j}u^{k}N^{i}N^{l}\right)\nonumber\\
&+\frac{\left(n^2-1-\mu_{\perp}\Delta\mu^{-1}\right)}{\mu_0\mu_{\perp}c^2}\left(g^{ik} u^{j} u^{l}-g^{il} u^{j} u^{k}+g^{jl} u^{i} u^{k}-g^{jk} u^{i} u^{l}\right)\nonumber\\
&+\frac{\Delta\mu^{-1}}{\mu_0}\left(g^{ik} N^{j} N^{l}-g^{il} N^{j} N^{k}+g^{jl} N^{i} N^{k}-g^{jk} N^{i} N^{l}\right),\label{explicitc}
\end{align}
where $\Delta\varepsilon:=\varepsilon_{||}-\varepsilon$ and $\Delta\mu^{-1}:=\mu^{-1}_{||}-\mu^{-1}$.

Applying the general procedure described in Sec.~\ref{dynamicalmediumm} to the Lagrangian of the total closed system with the anisotropic part (\ref{Lfluid1}) and constitutive tensor (\ref{explicitc}), after a rather long but straightforward calculation \cite{ORR02}, we obtain the following expression for the total BR tensor of the closed system:
\begin{align}
{\stackrel {\rm t}\sigma}_i{}^j={}&{\cal P}_{(i}u^{j)}-P^j_i\tilde{p}+{\stackrel F T}_{(i}{}^{j)}+\partial_k\left(J\nu u_i N^{[j}P^{k]}{}_{m}\dot{N}^{m}+J\nu u^j N_{[i}P^{k]}{}_{m}\dot{N}^{m}+N_{[i}\frac{\partial {\cal V}}{\partial(\partial_j N_{k]})}+N^{[j}\frac{\partial {\cal V}}{\partial(\partial^i N_{k]})}\right)\nonumber\\
{}&+\frac{\left(\mu^{-1} + \Delta\mu^{-1}\right)}{\mu_0}
\left[-\,F^{jk}F_{ik} + {\frac 14}\delta_i^j F^{kl}F_{kl}\right]+\,{\frac {\left(\varepsilon - \mu^{-1} - \Delta\mu^{-1}\right)}{\mu_0 c^2}}\left[
-\,F^{jk}u_k F_{il}u^l+\left(\frac{\delta_i^j}{2}-\frac{u_i u^j}{c^2}\right) (F_{kl}u^l)^2\right]\nonumber\\
{}&+\,\frac{1}{\mu_0}\Delta\mu^{-1}\left[-\,F^{jk}N_k F_{il}N^l + {\frac 12}
\delta_i^j(F_{kl}N^l)^2+N_{(i} F^{j)k}F_{kl}N^l-\frac{1}{c^2}N_{(i}u^{j)}F_{lk}u^kF^{lm}N_m\right]+\frac{1}{c^2}N_{(i}u^{j)}u_k\phi^k\nonumber\\
{}& +\,{\frac 1{\mu_0 c^2}}\left(\Delta\varepsilon + \Delta\mu^{-1}\right)
(F_{pq}N^p u^q)\left[\left(u_i u^j/c^2-\delta_i^j/2\right)(F_{kl}N^k u^l)+N_{(i} F^{j)k}u_k\right].\label{TBT}
\end{align}
Here we introduced the following quantities:
\begin{align}
{\cal P}_{i}:={}&\frac{u_i}{c^2}\left(\rho -\frac{1}{2}J\nu\omega^i\omega_i\right)-P_i{}^k\left[{\frac {J\nu}{c^2}}\dot{N}_k\,u^l\dot{N}_l-{\frac {\partial{\cal V}}{\partial u^k}}\right],\label{Phat}\\
\tilde{p}:={}&p+\,{\frac \nu2}\left(
\varepsilon_0\,{\frac {\partial\varepsilon}{\partial\nu}}\,{\cal E}^2 + {\frac 
1{\mu_0\mu_{\perp}^2}}\,{\frac {\partial\mu_{\perp}}{\partial\nu}}\,{\cal B}^2\right)-{\frac \nu2}\left(\varepsilon_0\,{\frac {\partial\Delta\varepsilon}
{\partial\nu}}\,({\cal E}_i N^i)^2 - {\frac 1{\mu_0}}\,{\frac {\partial
\Delta\mu^{-1}}{\partial\nu}}\,({\cal B}_i N^i)^2\right),\\
{\stackrel F T}_{i}{}^{j}:={}& -\,{\frac {\partial{\cal V}}{\partial
\partial_j N^k}}\partial_i N^k + \delta_i^j\,{\cal V},\\
\phi_i :={}& \partial_k\left(J\nu u^k P_i{}^j\dot{N}_j\right) 
- {\frac {\delta{\cal V}}{\delta N^i}}.
\end{align}
Notice that the abbreviation $(F_{kl}N^l)^2 = F_{kl}N^lF^{kn}N_n$ was used and that the total BR tensor (\ref{TBT}) is explicitly gauge invariant and symmetric. By inserting Eq.\,(\ref{explicitc}) in Eq.\,(\ref{4const}), one can check that the electromagnetic excitation $H^{ij}$ reads \cite{ORR02} explicitly
\begin{align}
H^{kl}={}&\frac{1}{\mu_0}\left(\mu^{-1}_{\perp} + \Delta\mu^{-1}\right)F^{kl}+\frac{2}{\mu_0}\Delta\mu^{-1}\,F^{[k}{}_n N^{l]}N^n+ {\frac {2}{\mu_0c^2}}\left(\varepsilon_{\perp} - \mu^{-1}_{\perp} - \Delta\mu^{-1}\right)F^{[k}{}_n u^{l]}u^n\nonumber\\
{}&- {\frac {2}{\mu_0c^2}}\left(\Delta\varepsilon + \Delta\mu^{-1}\right)N^{[k}u^{l]}\,F_{pq}N^p u^q.\label{Hab}
\end{align}
Then, substituting the expression above into the definition of the Abraham tensor (\ref{emtA}), we find explicitly
\begin{align}
\Omega{}_i{}^j=&\frac{\left(\mu^{-1} + \Delta\mu^{-1}\right)}{\mu_0}
\left[-\,F^{jk}F_{ik} + {\frac 14}\delta_i^j F^{kl}F_{kl}\right]+\,{\frac {\left(\varepsilon - \mu^{-1} - \Delta\mu^{-1}\right)}{\mu_0c^2}}\left[
-\,F^{jk}u_k F_{il}u^l+\left(\delta_i^j/2-u_iu^j/c^2\right) (F_{kl}u^l)^2\right]\nonumber\\
& +\,\frac{1}{\mu_0}\Delta\mu^{-1}\left[-\,F^{jk}N_k F_{il}N^l + {\frac 12}
\delta_i^j (F_{kl}N^l)^2-\,{\frac {1}{c^2}}N_{(i} u^{j)}F_{kp}u^pF^{kq}N_q+N_{(i} F^{j)k}F_{kl}N^l)\right]\nonumber\\
{}& +\,{\frac 1{\mu_0 c^2}}\left(\Delta\varepsilon + \Delta\mu^{-1}\right)
(F_{pq}N^p u^q)\left[\left(u_iu^j/c^2-\delta_i^j/2\right)(F_{kl}N^k u^l)+N_{(i} F^{j)k}u_k\right].\label{TA}
\end{align}
Finally, we calculate the kinetic energy-momentum tensor ${\stackrel {\rm m}\kappa}{}_i{}^j:={\stackrel {\rm t}\sigma}_i{}^j-\Omega_i{}^j$ for the liquid crystal, which reads
\begin{align}
{\stackrel {\rm m}\kappa}{}_i{}^j:={}&{\cal P}_{(i}u^{j)}-P^j_i\tilde{p}+{\stackrel F T}_{(i}{}^{j)}+\frac{1}{c^2}N_{(i}u^{j)}u_k\phi^k\nonumber\\
&+\partial_k\left(J\nu u_i N^{[j}P^{k]}{}_{m}\dot{N}^{m}+J\nu u^j N_{[i}P^{k]}{}_{m}\dot{N}^{m}+N_{[i}\frac{\partial {\cal V}}{\partial(\partial_j N_{k]})}+N^{[j}\frac{\partial {\cal V}}{\partial(\partial^i N_{k]})}\right).\label{kineticliquidcrystal}
\end{align}
\end{widetext}
Since Eq.\,(\ref{kineticliquidcrystal}) depends only on the material dynamical fields of the system, we verify that the general relation (\ref{sigmatot}) is indeed satisfied in this nontrivial case of a liquid crystal medium. By comparing Eq.\,(\ref{TA}) with the total BR tensor (\ref{TBT}), we see the nice result that all terms depending explicitly on the field strength are exactly those contained in the general algebraic definition (\ref{abraham}) of the Abraham tensor, in the same way as it was shown to happen for the isotropic case \cite{obukhov1}. By introducing the total BR tensor and using its relation (\ref{sigmatot}) to the Abraham tensor, we thus extend the previous result to general linear media.


\begin{thebibliography}{999}
\bibitem{abraham1} 
M. Abraham, {\it Zur Elektrodynamik bewegter K\"orper}, Rend. Circ. Mat. Palermo {\bf 28} (1909) 1-28.
\bibitem{abraham2} 
M. Abraham, {\it Sull'elettrodinamica di Minkowski}, Rend. Circ. Mat. Palermo {\bf 30} (1910) 33-46.
\bibitem{minkowski} 
H. Minkowski, {\it Die Grundgleichungen f\"{u}r die elektromagnetischen Vorg\"{a}nge in bewegten K\"{o}rpern}, Nachr. Ges.Wiss.G\"{o}tt. (1908) 53-111; reprinted in: Math. Ann. {\bf 68} (1910) 472-525 and in: {\it Zwei Abhandlungen ber die Grundgleichungen der Elektrodynamik} (B.G. Teubner Verlag, Leipzig und Berlin, 1910).
\bibitem{milonniboyd} 
P. Milonni and R. Boyd, {\it Momentum of light in a dielectric medium}, 
Adv. Opt. Phot. {\bf 2} (2010) 519-553.
\bibitem{dielectricslab} 
T. Ramos, G.F. Rubilar, and Yu.N. Obukhov, {\it Relativistic analysis of the dielectric Einstein box: Abraham, Minkowski and total energy-momentum tensors}, Phys. Lett. A {\bf 375} (2011) 1703-1709.
% http://dx.doi.org/10.1016/j.physleta.2011.03.015
\bibitem{pauli}
W. Pauli, {\it Relativit\"{a}tstheorie}, in: {\sl Encyklop\"{a}die der mathematischen Wissenschaften, Bd. V, Heft IV, Art. 19} (Teubner: Leipzig, 1921); W. Pauli, {\it Theory of relativity}, 2nd Edition (Pergamon Press, London, 1958).
\bibitem{wgordon}
W. Gordon, {\it Zur Lichtfortpflanzung nach der Relativit\"atstheorie.}, Ann. Phys. (Leipzig) {\bf 72} (1923) 421-456.
\bibitem{tamm1939}
I.E. Tamm, {\it Radiation emitted by uniformly moving electrons}, J. Phys. (Moscow) {\bf 1}, n. 5-6 (1939) 439-454.
\bibitem{vonlaue}
M. von Laue, {\it Zur Minkowskischen Elektrodynamik der bewegten K\"{o}rper}, Z. Phys. {\bf 128} (1950) 387-394.
\bibitem{moller1}
C. M\o ller, {\it The theory of relativity}, 1st ed. (Clarendon Press: Oxford, 1952), pages 202-211.
\bibitem{balazs}
N.L. Balazs, {\it The energy-momentum tensor of the electromagnetic field inside matter}, Phys. Rev. {\bf 91}  (1953) 408-411.
\bibitem{jpgordon} 
J.P. Gordon, {\it Radiation forces and momenta in dielectric media}, Phys. Rev. A {\bf 8} (1973) 14-21.
\bibitem{israel} 
W. Israel, {\it Relativistic effects in dielectrics: An experimental decision between Abraham and Minkowski?}, Phys. Lett. B {\bf 67} (1977) 125-128.
\bibitem{brevik} 
I. Brevik, {\it Experiments in phenomenological electrodynamics and the electromagnetic energy-momentum tensor}, Phys. Rep. {\bf 52} (1979) 133-201.
\bibitem{jonesrichards} 
R.V. Jones and J.C.S. Richards, {\it Pressure of radiation in a refracting medium}, Proc. Roy. Soc. Lond. A {\bf 221} (1954) 480-498.
\bibitem{james} 
R.P. James, {\it Force on permeable matter in time-varying fields}, 
Ph.D. thesis (Dept. of Electrical Engineering, Stanford University, 1968).
\bibitem{ashkin} 
A. Ashkin and J. Dziedzic, {\it Radiation pressure on a free liquid surface}, Phys. Rev. Lett. {\bf 30} (1973) 139-142.
\bibitem{wlw} 
G.B. Walker, D.G. Lahoz, and G. Walker, {\it Measurement of the Abraham force in a barium titanate specimen}, Can. J. Phys. {\bf 53} (1975) 2577-2586.
\bibitem{ww} 
G.B. Walker and G. Walker, {\it Mechanical forces in a dielectric due to electromagnetic fields}, Can. J. Phys. {\bf 55} (1977) 2121-2127.
\bibitem{jonesleslie} 
R.V. Jones and B. Leslie, {\it The measurement of optical radiation pressure in dispersive media}, Proc. Roy. Soc. Lond. A {\bf 360} (1978) 347-363.
\bibitem{gibson} 
A.F. Gibson, M.F. Kimmitt, A.O. Koohian, D.E. Evans, and G.F.D. Levy, {\it A study of radiation pressure in a refractive medium by the photon drag effect}, Proc. Roy. Soc. Lond. A {\bf 370} (1980) 303-311.
\bibitem{kristensen} 
M. Kristensen and J.P. Woerdman, {\it Is photon angular momentum conserved in a dielectric medium?}, Phys. Rev. Lett. {\bf 72} (1994) 2171-2174.
\bibitem{penfieldhaus1} 
P. Penfield and H.A. Haus, {\it Hamilton's principle for electromagnetic fluids}, Phys. Fluids {\bf 9} (1966) 1195-1204.
\bibitem{penfieldhaus1967} 
P. Penfield and H.A. Haus, {\it Electrodynamics of moving media} (MIT, Cambridge, MA, 1967).
\bibitem{robinson} 
F.N.H. Robinson, {\it Electromagnetic stress and momentum in matter}, Phys. Rep. {\bf 16} (1975) 313-354.
\bibitem{mikura} 
Z. Mikura, {\it Variational formulation of the electrodynamics of fluids and its application to the radiation pressure problem}, Phys. Rev. A {\bf 13} (1976) 2265-2275.
\bibitem{maugin} 
G.A. Maugin, {\it On the covariant equations of the relativistic electrodynamics of continua. I. General Equations}, J. Math. Phys. {\bf 19} (1978) 1198-1205.
\bibitem{kranys} 
M. Kranys, {\it The Minkowski and Abraham tensors, and the non-uniqueness of non-closed systems}, Int. J. Engng. Sci. {\bf 20} (1982) 1193-1213.
\bibitem{ashkintweesers} 
A. Ashkin, J. M. Dziedzic, J. E. Bjorkholm, and S. Chu, {\it Observation of a single-beam gradient force optical trap for dielectric particles}, Opt. Lett. {\bf 11} (1986) 288-290.
\bibitem{revolutiontweesers} 
D.G. Grier, {\it A revolution in optical manipulation}, Nature {\bf 424} (2003) 810-816.
% http://dx.doi.org/10.1038/nature01935
\bibitem{binding} 
K. Dholakia and P. Zem\'anek, {\it Gripped by light: Optical binding}, Rev. Mod. Phys. {\bf 82} (2010) 1767-1791.
% http://dx.doi.org/10.1103/RevModPhys.82.1767
\bibitem{cloak0} 
A. Greenleaf, Y. Kurylev, M. Lassas, U. Leonhardt, and G. Uhlmann, {\it Cloaked electromagnetic, acoustic, and quantum amplifiers via transformation optics}, Proc. Nat. Acad. Sci. {\bf 109} (2012) 10169-10174.
% http://dx.doi.org/10.1073/pnas.1116864109 
\bibitem{cloak}
X. Chen, Y. Luo, J. Zhang, K. Jiang, J.B. Pendry, and S. Zhang, {\it Macroscopic invisibility cloaking of visible light}, Nature Commun. {\bf 2} (2011) 176 (6 pages).
% http://dx.doi.org/10.1038/ncomms1176
\bibitem{loudon1} 
R. Loudon, {\it Theory of the radiation pressure on dielectric surfaces}, J. Mod. Opt. {\bf 49} (2002) 821-838.
\bibitem{obukhovhehl} 
Yu.N. Obukhov and F.W. Hehl, {\it Electromagnetic energy-momentum and forces in matter}, Phys. Lett. A {\bf 311} (2003) 277-284.
% http://dx.doi.org/10.1016/S0375-9601(03)00503-6
\bibitem{padgettbarnettloudon} 
M. Padgett, S. Barnett, and R. Loudon, {\it The angular momentum of light inside a dielectric}, J. Mod. Opt. {\bf 50} (2003) 1555-1562.
% http://dx.doi.org/10.1080/09500340308235229
\bibitem{loudon} 
R. Loudon, {\it Radiation pressure and momentum in dielectrics}, Fortschr. Phys. {\bf 52} (2004) 1134-1140.
\bibitem{garrison} 
J.C. Garrison and R.Y. Chiao, {\it Canonical and kinetic forms of the electromagnetic momentum in an ad hoc quantization scheme for a dispersive dielectric}, Phys. Rev. A {\bf 70} (2004) 053826 (8 pages).
\bibitem{milonniboyd2} 
P.W. Milonni and R.W. Boyd, {\it Recoil and photon momentum in a dielectric}, Laser Physics {\bf 15} (2005) 1432-1438.
\bibitem{leonhardt} 
U. Leonhardt, {\it Energy-momentum balance in quantum dielectrics}, Phys. Rev. A {\bf 73} (2006) 032108 (5 pages).
\bibitem{hindsbarnett}
E.A. Hinds and S.M. Barnett, {\it Momentum exchange between light and a single atom: Abraham or Minkowski?}, Phys. Rev. Lett. {\bf 102} (2009) 050403 (4 pages).
\bibitem{mansuripur5} 
M. Mansuripur and A.R. Zakharian, {\it Theoretical analysis of the force on the end face of a nanofilament exerted by an outgoing light pulse}, Phys. Rev. A {\bf 80} (2009) 023823 (7 pages).
\bibitem{mansuripur} 
M. Mansuripur, {\it Resolution of the Abraham-Minkowski controversy}, Opt. Comm. {\bf 283} (2010) 1997-2005.
\bibitem{crenshaw} 
M.E. Crenshaw and T.B. Bahder, {\it Energy-momentum tensor of the electromagnetic field in a dielectric}, Opt. Comm. {\bf 284} (2011) 2460-2465.
\bibitem{bradshawboydmilonni} 
D.H. Bradshaw, S. Shi, R.W. Boyd, and P.W. Milonni, {\it Electromagnetic momenta and forces in dispersive media}, Opt. Comm. {\bf 283} (2010) 650-656.
% http://dx.doi.org/10.1016/j.optcom.2009.10.056
\bibitem{shevchenko2} 
A. Shevchenko and M. Kaivola, {\it Electromagnetic force density in dissipative isotropic media}, J. Phys. B {\bf 44} (2011) 065403 (5 pages).
% http://dx.doi.org/10.1088/0953-4075/44/6/065403
\bibitem{brevik2010} 
I. Brevik and S.A. Ellingsen, {\it Possibility of measuring the Abraham force using whispering gallery modes}, Phys. Rev. A {\bf 81} (2010) 063830 (6 pages).
% http://dx.doi.org/10.1103/PhysRevA.81.063830
\bibitem{brevik2012} 
I. Brevik and S.A. Ellingsen, {\it Detection of the Abraham force with a succession of short optical pulses}, Phys. Rev. A {\bf 86} (2012) 025801 (4 pages).
\bibitem{kempletter} 
B.A. Kemp and T.M. Grzegorczyk, {\it The observable pressure of light in dielectric fluids}, Opt. Lett. {\bf 36} (2011) 493-495.
\bibitem{campbell} 
G.K. Campbell, A.E. Leanhardt, J. Mun, M. Boyd, E.W. Streed, W. Ketterle, and D.E. Pritchard, {\it Photon recoil momentum in dispersive media}, Phys. Rev. Lett. {\bf 94} (2005) 170403 (4 pages).
\bibitem{she} 
W. She, R. Yu, and R. Feng, {\it Observation of a push force on the end face of a nanometer silica filament exerted by outgoing light}, Phys. Rev. Lett. {\bf 101} (2008) 243601 (4 pages).
\bibitem{rikken} 
G.L. Rikken and B.A. van Tiggelen, {\it Measurement of the Abraham force and its predicted QED corrections in crossed electric and magnetic fields}, {\sl Phys. Rev. Lett.} {\bf 107} (2011) 170401 (4 pages).
\bibitem{rikken2} 
G.L.J.A. Rikken and B.A. van Tiggelen, {\it Observation of the intrinsic Abraham force in time-varying magnetic and electric fields}, Phys. Rev. Lett. {\bf 108} (2012) 230402 (4 pages).
\bibitem{pfeifer}
R.N.C. Pfeifer, T.A. Nieminen, N.R. Heckenberg, and H.~Rubinsztein-Dunlop, {\it Momentum of an electromagnetic wave in dielectric media}, Rev. Mod. Phys. {\bf 79} (2007) 830-851.
\bibitem{obukhov1} 
Yu.N. Obukhov, {\it Electromagnetic energy and momentum in moving media}, Ann. Phys. (Berlin) {\bf 17} (2008) 830-851.
% http://dx.doi.org/10.1002/andp.200810313
\bibitem{barnett2010} 
S. Barnett, {\it Resolution of the Abraham-Minkowski dilemma}, {\sl Phys. Rev. Lett.} {\bf 104} (2010) 070401 (4 pages).
% http://dx.doi.org/10.1103/PhysRevLett.104.070401
\bibitem{barnettloudon2} 
S. Barnett and R. Loudon, {\it The enigma of optical momentum in a medium}, Phil. Trans. R. Soc. A {\bf 368} (2010) 927-939.
% http://dx.doi.org/10.1098/rsta.2009.0207
\bibitem{saldanha} 
P.L. Saldanha, {\it Division of the momentum of electromagnetic waves in linear media into electromagnetic and material parts}, Optics Express {\bf 18} (2010) 2258-2268.
\bibitem{dodin} 
I.Y. Dodin and N. J. Fisch, {\it Axiomatic geometrical optics, Abraham-Minkowski controversy, and photon properties derived classically}, Phys. Rev. A {\bf 86} (2012) 053834 (16 pages).
% http://dx.doi.org/10.1103/PhysRevA.86.053834
\bibitem{crenshaw2013} 
M.E. Crenshaw, {\it Decomposition of the total momentum in a linear dielectric into field and matter components}, Ann. Phys. (N.Y.) {\bf 338} (2013) 97-106.
% http://dx.doi.org/10.1016/j.aop.2013.07.005
\bibitem{crenshaw2014} 
M.E. Crenshaw, {\it Electromagnetic momentum and the energy-momentum tensor in a linear medium with magnetic and dielectric properties}, J. Math. Phys. {\bf 55} (2014) 042901 (12 pages).
% http://dx.doi.org/10.1063/1.4869746
\bibitem{webb} 
K.J. Webb, {\it Dependence of the radiation pressure on the background refractive index}, Phys. Rev. Lett. {\bf 111} (2013) 043602 (5 pages).
\bibitem{kempresolution} 
B.A. Kemp, {\it Resolution of the Abraham-Minkowski debate: Implications for the electromagnetic wave theory of light in matter}, J. Appl. Phys. {\bf 109} (2011) 111101 (17 pages).
% http://dx.doi.org/10.1063/1.3582151
\bibitem{baxterloudon} 
C. Baxter and R. Loudon, {\it Tutorial review: Radiation pressure and the photon momentum in dielectrics}, J. Mod. Opt. {\bf 57} (2010) 830-842.
\bibitem{griffiths} 
D.J. Griffiths, {\it Resource letter EM-1: Electromagnetic momentum}, Am. J. Phys. {\bf 80} (2012) 7-18.
% http://dx.doi.org/10.1119/1.3641979
\bibitem{ORR02} 
Yu.N. Obukhov, T. Ramos, and G.F. Rubilar, {\it Relativistic Lagrangian model of a nematic liquid crystal interacting with an electromagnetic field}, Phys. Rev. E {\bf 86} (2012) 031703 (18 pages).
% http://dx.doi.org/10.1103/PhysRevE.86.031703
\bibitem{obukhovhehl2007} 
Yu.N. Obukhov and F.W. Hehl, {\it Electrodynamics of moving magnetoelectric media: Variational approach}, Phys. Lett. A {\bf 371} (2007) 11-19.
% http://dx.doi.org/10.1016/j.physleta.2007.08.026
\bibitem{Birk} 
F.W. Hehl and Yu.N. Obukhov, {\it Foundations of classical electrodynamics -- Charge, flux, and metric} (Birkh\"auser, Boston, MA, 2003).
\bibitem{footnote} Notice that here we use a slightly different convention for the Lorentz generators as compared to $(\rho_{kl})^A{}_B$ in \cite{obukhov1,ORR02}, they are related via $(\rho_{kl})^A{}_B=-(s_{kl})^A{}_B/2$.
\bibitem{hehl76}
F.W. Hehl, {\it On the energy tensor of spinning massive matter in classical field theory}, Rept. Math. Phys. {\bf 9} (1976) 55-82.
\bibitem{Schueking}
F.W. Hehl, A. Mac\'{\i}as, E.W. Mielke, and Yu.N. Obukhov, {\it On the structure of the energy-momentum and the spin currents in Dirac's electron theory}, in: {\sl ``On Einstein's path", Essays in honor of E.~Schucking}, Ed. A. Harvey (Springer: New York, 1998) 257-274; arxiv:gr-qc/9706009.
\bibitem{Belinfante1}
F.~J.~Belinfante, {\it On the spin angular momentum of mesons}, Physica {\bf 6} (1939) 887-898.
\bibitem{Belinfante2}
F.~J.~Belinfante, {\it On the current and the density of the electric charge, the energy, the linear momentum and the angular momentum of arbitrary fields}, Physica {\bf 7} (1940) 449-474.
\bibitem{Rosenfeld} 
L.~Rosenfeld, {\it Sur le tenseur d'impulsion-\'energie}, M\'em. Acad. Roy. Belgique, cl. sc. {\bf 18}, fasc. 6 (1940) 1-30; English translation: L.~Rosenfeld, {\it On the energy-momentum tensor}, in: {\sl ``Selected Papers of L\'eon Rosenfeld"}, Eds. R.S. Cohen and J.J. Stachel (D. Reidel Publishing Company, Dordrecht, 1979) pp. 711-735.
\bibitem{Bliokh2014}
K.Y. Bliokh, A.Y. Bekshaev, and F. Nori, {\it Extraordinary momentum and spin in evanescent waves}, Nature Commun. {\bf 5} (2014) 3300 (29 pages).
% http://dx.doi.org/10.1038/ncomms4300
\bibitem{landau} 
L.D. Landau and E.M. Lifshitz, {\it The classical theory of fields}, 4th ed. (Elsevier Butterworth-Heinemann, Oxford, 2000).
\bibitem{jackson} 
J.D. Jackson, {\it Classical electrodynamics}, 3rd ed. (John Wiley \&  Sons, Inc., 1999).
\bibitem{MTW} 
C.W. Misner, K.S. Thorne, and J.A. Wheeler, {\it Gravitation} (W.H. Freeman and Company, San Francisco, 1973).
\bibitem{kundu}
P.K. Kundu and I.M. Cohen, {\it Fluid Mechanics}, 4th ed. (Elsevier: Oxford, 2008).
\bibitem{TT}
C. Truesdell and R.A. Toupin, {\it The classical field theories}, in: {\sl ``Principles of classical mechanics and field theory''}, Handbuch der Physik, Ed. S. Fl\"ugge (Springer: Berlin, 1960) vol. III/1, pp. 226-793.
\bibitem{ann2012}
J. Gratus, Yu.N. Obukhov, and R.W. Tucker, {\it Conservation laws and stress-energy-momentum tensors for systems with background fields}, Ann. Phys. (N.Y.) {\bf 327} (2012) 2560-2590.
\bibitem{LL}
L.D. Landau and E.M. Lifshitz, {\it Electrodynamics of continuous media}, 2nd ed. (Elsevier Butterworth-Heinemann, Oxford, 2004).
\bibitem{Post}
E.J. Post, {\it Formal structure of electromagnetics -- General covariance and electromagnetics} (North Holland, Amsterdam, 1962) and (Dover, Mineola, NY, 1997).
\bibitem{Furutsu}
K. Furutsu, {\it Energy-momentum tensor of electromagnetic field in moving dispersive media and instability: relativistic formulation}, Phys. Rev. {\bf 185} (1969) 257-272.
\bibitem{Lindell}
I.V. Lindell, A.H. Sihvola, S.A. Tretyakov, and A.J. Viitanen, {\it Electromagnetic waves in chiral and bi-isotropic media} (Artech House, Boston, 1994).
\bibitem{itin} 
Y. Itin, {\it Dispersion relation for electromagnetic waves in anisotropic media}, Phys. Lett. A {\bf 374} (2010) 1113-1116.
% http://dx.doi.org/10.1016/j.physleta.2009.12.071
\bibitem{favaro} 
I. Lindell, L. Bergamin, and A. Favaro, {\it Decomposable medium conditions in four-dimensional representation}, IEEE Trans. Antennas and Propagation {\bf 60} (2012) 367-376.
% http://dx.doi.org/10.1109/TAP.2011.2167937
\bibitem{dahl} 
M.F. Dahl, {\it Non-dissipative electromagnetic media with two Lorentz null cones}, Ann. Phys. (N.Y.) {\bf 330} (2013) 55-73.
% http://dx.doi.org/10.1016/j.aop.2012.11.005
\bibitem{Odell}
T.H. O'Dell, {\it The electrodynamics of magneto-electric media} (North-Holland, Amsterdam, 1970).
\bibitem{Bateman}
H. Bateman, {\it Kummer's quartic surface as a wave surface},
Proc. London Math. Soc. {\bf 8}(1) (1910) 375-382.
\bibitem{Tamm1} 
I.E. Tamm, {\it Relativistic crystal optics and its relation to the 
geometry of a bi-quadratic form}, Zhurn. Ross. Fiz.-Khim. Ob. 
{\bf 57}, n. 3-4 (1925) 209-224 (in Russian); Reprinted in: I.E. Tamm, {\it Collected Papers} (Nauka: Moscow, 1975) Vol.\ 1, pp.\ 33-61 (in Russian).
\bibitem{Tamm2} 
I.E. Tamm, {\it Electrodynamics of an anisotropic medium in special 
relativity theory}, Zhurn. Ross. Fiz.-Khim. Ob. {\bf 56}, 
n. 2-3 (1924) 248-262 (in Russian); Reprinted in: I.E. Tamm, 
{\it Collected Papers} (Nauka: Moscow, 1975) Vol.\ 1, pp.\ 19-32 (in Russian). 
\bibitem{Tamm3} 
L. Mandelstam and J. Tamm, {\it Elektrodynamik der anisotropen Medien in der speziellen Relativit\"atstheorie}, Math. Ann. {\bf 95} (1925) 154-160; Reprinted in: I.E. Tamm, {\it Collected Papers} (Nauka: Moscow, 1975) Vol.\ 1, pp.\ 62-67 (in Russian).
\bibitem{lin}
C.C. Lin, {\it Hydrodynamics of Helium II}, in: {\sl Liquid Helium}, Proc. of XXI Internat. School of Physics ``Enrico Fermi'', Varenna 3-15 July 1961, ed. G. Careri (Academinc Press: New York, 1963) pp. 93-164. 
\bibitem{ray72} 
J.R. Ray, {\it Lagrangian density for perfect fluids in general relativity},
J. Math Phys. {\bf 13} (1972) 1451-1453.
\bibitem{WR}
J. Weyssenhoff and A. Raabe, {\it Relativistic dynamics of spin-fluids
and spin-particles}, Acta Phys. Pol. {\bf 9} (1947) 7-18.
\bibitem{halb}
F. Halbwachs, {\it Lagrangian formalism for a classical relativistic
particle endowed with internal structure}, Progr. Theor. Phys. {\bf 24} (1960) 291-307.
\bibitem{Kopc}
W. Kopczy\'nski, {\it Lagrangian dynamics of particles and fluids with
intrinsic spin in Einstein-Cartan space-time}, Phys. Rev. D {\bf 34} (1986) 352-356.
\bibitem{OK}
Yu.N. Obukhov and V.A. Korotky, {\it The Weyssenhoff fluid in Einstein-Cartan theory}, Class. Quantum Grav. {\bf 4} (1987) 1633-1657.
% http://dx.doi.org/10.1088/0264-9381/4/6/021
\bibitem{OT}
Yu.N. Obukhov and R. Tresguerres, {\it Hyperfluid - a model of classical matter with hypermomentum}, Phys. Lett. A {\bf 184} (1993) 17-22.
% http://dx.doi.org/10.1016/0375-9601(93)90339-2 
\bibitem{fukuma1} 
M. Fukuma and Y. Sakatani, {\it Entropic formulation of relativistic continuum mechanics}, Phys. Rev. E {\bf 84} (2011) 026315 (13 pages).
% http://dx.doi.org/10.1103/PhysRevE.84.026315
\bibitem{fukuma2} 
M. Fukuma and Y. Sakatani, {\it Relativistic viscoelastic fluid mechanics}, Phys. Rev. E {\bf 84} (2011) 026316 (25 pages).
% http://dx.doi.org/10.1103/PhysRevE.84.026316
\bibitem{philbin}
T.G. Philbin, {\it Electromagnetic energy momentum in dispersive media}, Phys. Rev. A {\bf 83} (2011) 013823 (7 pages).
% http://dx.doi.org/10.1103/PhysRevA.83.013823
\bibitem{philbin2} 
T.G. Philbin and O. Allanson, {\it Optical angular momentum in dispersive media}, Phys. Rev. A {\bf 86} (2012) 055802 (4 pages).
% http://dx.doi.org/10.1103/PhysRevA.86.055802
\bibitem{mashhoon1} 
B. Mashhoon, {\it Vacuum electrodynamics of accelerated systems: Nonlocal Maxwell's equations}, Ann. Phys. (Leipzig) {\bf 12} (2003) 586-598.
% http://dx.doi.org/10.1002/andp.200310028
\bibitem{mashhoon2} 
B. Mashhoon, {\it Nonlocal electrodynamics of rotating systems}, Phys. Rev. A {\bf 72} (2005) 052105 (10 pages).
% http://dx.doi.org/10.1103/PhysRevA.72.052105
\bibitem{mashhoon3} 
B. Mashhoon, {\it Nonlocal electrodynamics of accelerated systems}, Phys. Lett. A {\bf 366} (2007) 545-549. 
% http://dx.doi.org/10.1016/j.physleta.2007.02.071
\bibitem{castaldi} 
G. Castaldi, V. Galdi, A. Alu, and N. Engheta, {\it Nonlocal transformation optics}, Phys. Rev. Lett. {\bf 108} (2012) 063902 (5 pages).
% http://dx.doi.org/10.1103/PhysRevLett.108.063902
\bibitem{ogr}
Yu.N. Obukhov and G.F. Rubilar, {\it Fresnel analysis of wave propagation in nonlinear electrodynamics}, Phys. Rev. D {\bf 66} (2002) 024042 (11 pages).
% http://dx.doi.org/10.1103/PhysRevD.66.024042
\bibitem{lorenci} 
V.A. De Lorenci, R. Klippert, and D.H. Teodoro, {\it Birefringence in nonlinear anisotropic dielectric media}, Phys. Rev. D {\bf 70} (2004) 124035 (5 pages).
% http://dx.doi.org/10.1103/PhysRevD.70.124035
\bibitem{pereiraklippert} 
D.D. Pereira and R. Klippert, {\it Local nonlinear electrodynamics}, Phys. Lett. A {\bf 374} (2010) 4175-4179.
% http://dx.doi.org/10.1016/j.physleta.2010.08.033
\end{thebibliography}
\end{document}